\def\compiletechreport{1}
\documentclass[runningheads]{llncs}
\usepackage{xspace}
\usepackage{graphicx}
\usepackage{psfrag}
\usepackage{enumitem} 
\usepackage{pifont}
\usepackage{color}    
\usepackage{threeparttable}
\usepackage{rotating}
\usepackage{colortbl} 
\usepackage{epsfig}
\usepackage{cite}
\usepackage{subfigure}
\usepackage[ruled]{algorithm2e}
\usepackage{multicol}
\usepackage{amsmath, amssymb}
\usepackage[noend]{algorithmic}
\usepackage{multirow}
\usepackage{hhline}
\usepackage[font={normal, it}]{caption}
\usepackage[normalem]{ulem}  
\usepackage{comment}

\usepackage{tabulary}
\usepackage{booktabs}
\usepackage{array}

\usepackage{hyperref}
\hypersetup{
    colorlinks=false,
    citecolor=black,
    linkcolor=black,
    filecolor=magenta,      
    urlcolor=cyan,
}

\newif\iftechreport
\ifdefined\compiletechreport
    \techreporttrue
\else
    \techreportfalse
\fi

\newcommand{\danielecolor}{magenta}
\newcommand{\danielesubscript}{\textsubscript{\textcolor{\danielecolor}{\textsf{\textbf{DA}}}}}

\newcommand{\dadelnoname}[1]{\bgroup\markoverwith{\textcolor{\danielecolor}{\rule[0.35ex]{2pt}{1pt}}}\ULon{#1}}
\newcommand{\dadel}[1]{\dadelnoname{#1}\kern0.1em\danielesubscript}

\newcounter{linecounter}

\newcommand{\etal}{et~al.\@\xspace}

\newcommand{\EFD}{RLFD\xspace}
\newcommand{\EFDs}{RLFDs\xspace}
\newcommand{\twinEFD}{Twin-RLFD\xspace}
\newcommand{\ETE}{CLEF\xspace}

\newcommand{\name}{CLEF\xspace}
\newcommand{\nameacr}{in-Core Limiting of Egregious Flows\xspace}

\renewcommand{\paragraph}[1]{\vspace{1.5mm}\noindent{\textbf{#1}}\quad}

\newcommand{\func}[2]{\mathsf{#1}(#2)}

\def\mit#1{\mbox{\it #1}}
\def\fmit#1{\mbox{\it \scriptsize #1}}
\def\ffmit#1{\mbox{\it \tiny #1}}

\def\centerhack#1{\hbox to 0pt{\hss\footnotesize #1\hss}}
\def\dchack#1{\vbox to 0pt{\vss{\hbox to 0pt{\hss#1\hss}}\vss}}

\begin{document}

\iftechreport
    \title{\name: Limiting the Damage Caused by Large Flows in the Internet Core (Technical Report)}
\else
    \title{\name: Limiting the Damage Caused by Large Flows in the Internet Core}
\fi

\author{
Hao Wu
\inst{1, 2}
\orcidID{0000-0002-5100-1519}
\and
Hsu-Chun Hsiao
\inst{3}
\orcidID{0000-0001-9592-6911}\and\\
Daniele E. Asoni
\inst{4}
\orcidID{0000-0001-5699-9237}\and\\
Simon Scherrer
\inst{4}
\orcidID{0000-0001-9557-1700}
\and
Adrian Perrig
\inst{4}
\orcidID{0000-0002-5280-5412}\and\\
Yih-Chun Hu
\inst{1}
\orcidID{0000-0002-7829-3929}
}

\authorrunning{
H. Wu,
H-C. Hsiao,
D. E. Asoni,
S. Scherrer,
A. Perrig 
and Y-C. Hu
}
\institute{
University of Illinois at Urbana Champaign
\and
Rubrik, Inc.
\and
National Taiwan University
\and
ETH Zurich
}

\maketitle

\begin{abstract}
The detection of network flows that send excessive amounts of traffic is of
increasing importance to enforce QoS and to counter DDoS attacks. Large-flow
detection has been previously explored, but the proposed approaches can be used
on high-capacity core routers only at the cost of significantly reduced
accuracy, due to their otherwise too high memory and processing overhead.
We propose \name, a new large-flow detection scheme with low memory
requirements, which maintains high accuracy under the strict conditions of
high-capacity core routers.
We compare our scheme with previous proposals through extensive theoretical
analysis, and with an evaluation based on worst-case-scenario attack traffic.
We show that \name outperforms previously proposed systems in settings with
limited memory.

\end{abstract}

\section{Introduction}
\label{sec:intro}

Detecting misbehaving large network flows\footnote{As in prior
literature~\cite{Estan2003,eardet}, the term \emph{large flow} denotes a flow
that sends more than its allocated bandwidth.} that use more than their
allocated resources is not only an important mechanism for Quality of Service
(QoS)~\cite{rfc2212} schemes such as IntServ~\cite{rfc1633}, but also for DDoS
defense mechanisms that allocate bandwidth to network
flows~\cite{BRSPZHKU2016,LiJiHuBa2016,LeKaGl2013}. With the recent resurgence of
volumetric DDoS attacks~\cite{Antonakakis2017}, the topics of DDoS defense
mechanisms and QoS are gaining importance; thus, the need for efficient
in-network accounting is increasing.

Unfortunately, per-flow resource accounting is too expensive to perform in the
core of the network~\cite{Estan2003}, since large-scale Internet core routers
have an aggregate capacity of several Terabits per second (Tbps). Instead, to
detect misbehaving flows, core routers need to employ highly efficient schemes
which do not require them to keep per-flow state.
Several approaches for large-flow detection have been proposed in this context;
they can be categorized into probabilistic (i.e., relying on random sampling or
random binning) and deterministic algorithms. Examples of probabilistic
algorithms are Sampled Netflow~\cite{Netflow} and Multistage
Filters~\cite{Estan2003, Estan2003a}, while EARDet~\cite{eardet} and Space
Saving~\cite{metwally2005efficient} are examples of deterministic approaches.

However, previously proposed algorithms are able to satisfy the requirements of
core router environments only by significantly sacrificing their accuracy. In
particular, with the constraints on the amount of high-speed memory on core
routers, these algorithms either can only detect flows which exceed their
assigned bandwidth by very large amounts, or else they suffer from high
false-positive rates. This means that these systems cannot prevent the
performance degradation of regular, well-behaved flows, because of large flows
that manage to stay ``under the radar'' of the detection algorithms, or because
the detection algorithms themselves erroneously flag and punish the well-behaved
flows.

As a numeric example, consider that for EARDet to accurately detect misbehaving
flows exceeding a threshold of 1~Mbps on a 100~Gbps link, it would require
$10^5$ counters for that link. Maintaining these counters, together with the
necessary associated metadata, requires between 1.6~MB and 4MB of
state\footnote{The IP metadata consists of source and destination addresses,
protocol number, and ports. Thus, it requires about 16~bytes and 40~bytes per
counter for IPv4 and IPv6, respectively.}, which exceeds typical high-speed
memory provisioning for core routers, and would come at a high cost (for
comparison, note that only the most high-end commodity CPUs approach the 1--4~MB
range with their per-core L1/L2 memory, and the price tag for such processors
surpasses USD 4000~\cite{intel-xeon-e7}).

In this paper we propose a novel randomized algorithm for large flow detection
called \emph{Recursive Large-Flow Detection} (\EFD). \EFD works by considering a
set of potential large flows, dividing this set into multiple subsets, and then
recursively narrowing down the focus to the most promising subset. This
algorithm is highly memory efficient, and is designed to have no false
positives. To achieve these properties, \EFD sacrifices some detection speed, in
particular for the case of multiple concurrent large flows. We improve on these
limitations by combining \EFD with the deterministic EARDet, proposing a hybrid
scheme called \name, short for \emph{\nameacr}. We show how this scheme
inherits the strengths of both algorithms: the ability to quickly detect very
large flows of EARDet (which it can do in a memory efficient way), and the
ability to detect low-rate large flows with minimal memory footprint of \EFD.

To have a significant comparison with related work, we define a \emph{damage}
metric which estimates the impact of failed, delayed, and incorrect detection on
well-behaved flows. We use this metric to compare \EFD and \name with previous
proposals, which we do both on a theoretical level and by evaluating the amount
of damage caused by (worst-case) attacks. Our evaluation shows that \name
performs better than previous work under realistic memory constraints, both in
terms of our damage metric and in terms of false negatives and false positives.

To summarize, this paper's main contributions are the following: a novel,
randomized algorithm, \EFD, that provides eventual detection of persistently
large flows with very little memory cost; a hybrid detection scheme, \name,
which offers excellent large-flow detection properties with low resource
requirements; the analysis of worst-case attacks against the proposed large-flow
detectors, using a damage metric that allows a realistic comparison with the
related work.

\section{Problem Definition}
\label{sec:def}

This paper aims to design an efficient large-flow detection algorithm
that minimizes the \emph{damage} caused by misbehaving flows. This
section introduces the challenges of large-flow detection and
defines a damage metric to compare different large-flow detectors. We then
define an adversary model in which the adversary adapts its behavior to the
detection algorithm in use.

\subsection{Large-Flow Detection}\label{ssec:large-flow-detection}

A flow is a collection of related traffic; 
for example, Internet flows are commonly characterized by a 5-tuple
(source / destination IP / port, transport protocol).
A \emph{large flow} is one that exceeds a flow specification
during a period of length $t$.
A flow specification can be defined using a leaky bucket
descriptor $\func{TH}{t} = \gamma t + \beta$,
where $\gamma > 0$ and $\beta > 0$ are the
maximum legitimate rate and burstiness allowance, respectively.
Flow specifications can be enforced in two ways:
\emph{arbitrary-window},
in which the flow specification is enforced over every possible
starting time,
or \emph{landmark-window},
in which the flow specification is enforced over
a limited set of starting times.

Detecting every large flow exactly when it exceeds the flow specification, and
doing so with no false positives requires per-flow state (this can be shown by
the pigeonhole principle~\cite{trybulec1990pigeon}),
which is expensive on core routers.
In this paper, we develop and evaluate schemes that trade timely detection
for space efficiency.

As in prior work in flow monitoring, we assume each flow has a unique and
unforgeable flow ID, e.g., using
source authentication techniques such as
accountable IPs~\cite{Andersen2008}, ICING~\cite{Naous2011},
IPA~\cite{LiLiYa2011}, OPT~\cite{kim2014lightweight},
or with Passport~\cite{Liu2008a}.
Such techniques can be deployed in the current Internet or in a future Internet
architecture, e.g., Nebula~\cite{anderson2013nebula},
SCION~\cite{zhang2011scion}, or XIA~\cite{xia}.

\paragraph{Large-flow detection by core routers.}
In this work, we aim to design a large-flow detection algorithm that is viable
to run on Internet core routers. The algorithm needs to limit damage caused by large flows
even when handling worst-case background traffic. Such an algorithm must satisfy
these three requirements:
\begin{itemize}
\item\textbf{Line rate:}
An in-core large-flow detection algorithm must operate at the line rate of core
routers, which can process several hundreds of gigabits of traffic per second.
\item \textbf{Low memory:}
Large-flow detection algorithms will typically access one or more memory
locations for each traversing packet; such memory must be high-speed (such as
on-chip L1 cache).
Additionally, such memory is expensive and usually limited in size,
and existing large-flow detectors are inadequate
to operate in high-bandwidth, low-memory environments.  An in-core
large-flow detection algorithm should thus be highly space-efficient.
Though perfect detection requires counters
equal to the maximum number of simultaneous large flows
(by the pigeonhole
principle~\cite{trybulec1990pigeon}),
our goal is to perform effective detection with much fewer counters.

\item \textbf{Low damage:}
With the performance constraints of the previous two points, the large-flow
detection algorithm should also minimize the damage to honest flows, which can
be caused either by the excessive bandwidth usage by large flows, or by the
erroneous classification of legitimate flows as large flows (false positives).
Section~\ref{ssec:def-metric} introduces our damage metric, which takes both
these aspects into account.

\end{itemize}

\subsection{Damage Metric}\label{ssec:def-metric}
\newcommand{\damage}{D}
\newcommand{\damageover}{\damage_{\fmit{over}}}
\newcommand{\damagefp}{\damage_{\fmit{fp}}}
We consider misbehaving large flows to be a problem mainly in that they have an
adverse impact on honest flows. To measure the performance of large flow
detection algorithms we therefore adopt a simple and effective \emph{damage}
metric which captures the packet loss suffered by honest flows. This metric
considers both (1) the direct impact of excessive bandwidth usage by large
flows, and (2) the potential adverse effect of the detection algorithm itself,
which may be prone to false positives resulting in the blacklisting of honest
flows. Specifically, we define our damage metric as
$\damage = \damageover + \damagefp$, where $\damageover$ (\emph{overuse damage})
is the total amount of traffic by which all large flows exceed the flow
specification, and $\damagefp$ (\emph{false positive damage}) is the amount of
legitimate traffic incorrectly blocked by the detection algorithm.
The definition of the overuse damage assumes a link at full capacity, so when
this is not the case the damage metric represents an over-approximation of the
actual traffic lost suffered by honest flows.
We note that the metrics commonly used by previous work, i.e., false positives,
false negatives, and detection delay, are all reflected by our metric.

\subsection{Attacker Model}\label{ssec:def-adv-model}
In our attacker model, we consider an adversary that aims to maximize damage.
Our attacker responds to the detection algorithm
and tries to exploit its transient behavior
to avoid detection or to cause false detection of legitimate flows.

Like Estan and Varghese's work~\cite{Estan2003},
we assume that attackers know about the large-flow detection
algorithm running in the router and its settings,
but have no knowledge of secret seeds used to generate random variables,
such as the detection
intervals for landmark-window-based algorithms~\cite{Misra1982, Demaine2002,
  Karp2003a, Manku2002, metwally2005efficient, Estan2003, Fang1999,
  Cormode2005},
and random numbers used for packet/flow sampling~\cite{Estan2003}.
This assumption prevents the attacker from performing optimal attacks against
randomized algorithms.

\begin{figure}[!th]
\centering
\includegraphics[width=0.9\columnwidth]{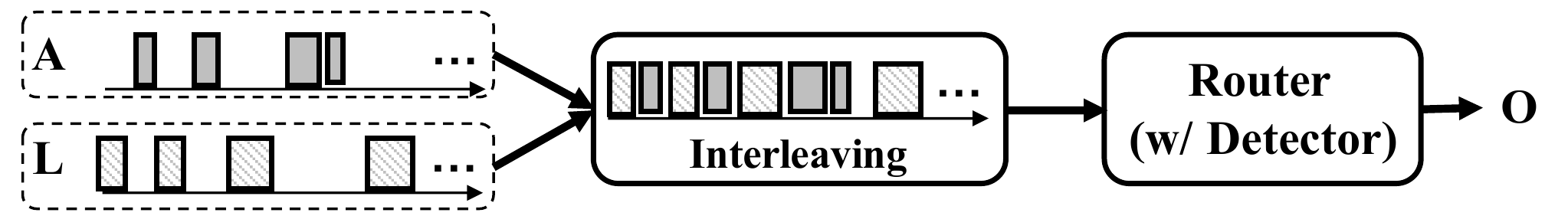}
\vspace*{-3mm}
\caption{Adversary Model.}
\label{fig:adv-model}
\end{figure}

We assume the attacker can interleave packets,
but is unable to spoof legitimate packets
(as discussed in Section~\ref{ssec:large-flow-detection})
or create pre-router losses in legitimate flows.
Figure~\ref{fig:adv-model} shows the network model,
where the attacker arbitrarily interleaves attack traffic ($A$)
between idle intervals of legitimate traffic ($L$),
and the router processes the interleaved traffic
to generate output traffic ($O$) and perform large-flow detection.
Our model does not limit input traffic,
allowing for arbitrary volumes of attack traffic.

In our model, whenever a packet traverses a router,
the large-flow detector receives the flow ID
(for example, the source and destination IP and port and transport protocol),
the packet size, and the timestamp at which the packet arrived.

\vspace{-3mm}
\vspace{-3mm}

\section{Background and Challenges}
\label{sec:overview}

In this section we briefly review some existing large flow detection
algorithms, and discuss the motivations and challenges
of combining multiple algorithms into a hybrid scheme.

\subsection{Existing Detection Algorithms}
\label{ssec:existing-algorithms}

We review the three most relevant large-flow detection algorithms,
 summarized in Table~\ref{table:detector-category}.
We divide large flows into \emph{low-rate large flows}
and \emph{high-rate large flows},
depending on the amount by which they exceed the flow specification.

\paragraph{EARDet.}
EARDet~\cite{eardet} guarantees exact and instant detection of all flows
exceeding a high-rate threshold $\gamma_h = \frac{\rho}{m}$,
where $\rho$ is the link capacity and $m$ is the number of counters. However,
EARDet may fail to identify a large flow whose rate stays below $\gamma_h$.

\paragraph{Multistage Filters.}
Multistage filters~\cite{Estan2003, Estan2003a} consist of multiple parallel
stages, each of which is an array of counters. Specifically,
\emph{arbitrary-window-based Multistage Filter} (AMF), as classified by
Wu~\etal~\cite{eardet}, uses leaky buckets as counters. AMF guarantees the
absence of false negatives (no-FN) and immediate detection for any flow
specification; however, AMF has false positives (FPs), which increase as the
link becomes
\iftechreport
congested (as shown in Appendix~\ref{ssec:appendix-mf-analysis}).
\else
congested.
\fi

\paragraph{Flow Memory.}
Flow Memory (FM)~\cite{Estan2003} refers to per-flow monitoring
of select flows.
FM is often used in conjunction with another system
that specifies which flows to monitor;
when a new flow is to be monitored but the flow memory is full,
FM evicts an old flow.
We follow Estan and Varghese~\cite{Estan2003}'s random eviction.
If the flow memory is large enough to avoid eviction,
it provides exact detection.
In practice, however,
Flow Memory is unable to handle a large number of flows,
resulting in frequent flow eviction
and potentially high FN.
\iftechreport
The analysis in Appendix~\ref{ssec:appendix-fm-analysis} shows that
\fi
FM's real-world performance depends on
the amount by which a large flow exceeds the flow specification:
high-rate flows are more quickly detected,
which improves the chance of detection before eviction.

\vspace{-3mm}
\begin{table}[htbp]\caption{Comparison of three existing detection algorithms.
    None of them achieve all desired properties.}
\label{table:detector-category}
\centering
\begin{small}
  \begin{tabular}{|c|c|c|c|c|}
    \hline
    \hline
    \multicolumn{2}{|c|}{Algorithm} & EARDet & AMF & FM\\
    \hline
    \multicolumn{2}{|c|}{No-FP} & yes & no$^{*}$ & yes \\
    \hline
    \multirow{2}{*}{No-FN} & low-rate  & no$^{**}$ & yes & no$^{*}$\\
    \hhline{~----}        & high-rate & yes       & yes & yes$^{***}$\\
    \hline
    \multicolumn{2}{|c|}{Instant detection} & yes & yes & yes \\
    \hline
  \end{tabular}

\iftechreport
  $^{*}$Appendix~\ref{ssec:appendix-fm-analysis}
  and~\ref{ssec:appendix-mf-analysis} show that Flow Memory has high FN and AMF
  has high FP for low-rate large flows when memory is limited.
\else
\vspace{2mm}
  $^{*}$In our technical report~\cite{full-paper} we show that Flow Memory has
  high FN and AMF has high FP for low-rate large flows when memory is limited.
\fi

  $^{**}$EARDet cannot provide no-FN when memory is limited.

  $^{***}$Flow Memory has nearly zero FN when large-flow rate is high.
\end{small}
\end{table}
\vspace{-3mm}

\subsection{Advantages of Hybrid Schemes}\label{ssec:why-hybrid}

As Table~\ref{table:detector-category} shows,
none of the detectors we examined can
efficiently achieve no-FN and no-FP
across various types of large flows.
However, different detectors exhibit different strengths,
so combining them could result in improved performance.

One approach is to run detectors sequentially;
in this composition, the first detector monitors all traffic
and sends any large flows it detects to a second detector.
However, this approach allows an attacker controlling multiple flows
to rotate overuse among many flows, overusing a flow only for as long as it
takes the first detector to react, then sending at the normal rate so that
remaining detectors remove it from their watch list and re-starting with
the attack.

Alternatively, we can run detectors in parallel: the hybrid detects a flow
whenever it is identified by either detector. (Another configuration is that a
flow is only detected if both detectors identify it, but such a configuration
would have a high FN rate compared to the detectors used in this paper.) The hybrid
inherits the FPs of both schemes, but features the minimum detection delay of
the two schemes and has a FN only when both schemes have a FN. The remainder of
this paper considers the parallel approach that identifies a flow whenever it is
detected by either detector.

The EARDet and Flow Memory schemes have no FPs
and are able to quickly detect high-rate flows;
because high-rate flows cause damage much more quickly,
rapid detection of high-rate flows is important to achieving low damage.
Combining EARDet or Flow Memory
with a scheme capable of detecting low-rate flows as a hybrid detection scheme
can retain rapid detection of high-rate flows
while eventually catching (and thus limiting the damage of) low-rate flows.
In this paper, we aim to construct such a scheme.
Specifically, our scheme will selectively monitor one small set at a time,
ensuring that a consistently-overusing flow is eventually detected.

\section{\EFD and \name Hybrid Schemes}
\label{sec:hybrid}

In this section, we present our new large-flow detectors. First, we describe the
\emph{Recursive Large-Flow Detection} (\EFD) algorithm, a novel approach which
is designed to use very little memory but provide eventual detection for large
flows. We then present the data structures, runtime analysis, and advantages and
disadvantages of \EFD. Next, we develop a hybrid detector, \name, that addresses
the disadvantages of \EFD by combining it with the previously proposed
EARDet~\cite{eardet}. \name uses EARDet to rapidly detect high-rate flows and
\EFD to detect low-rate flows, thus limiting the damage caused by large flows,
even with a very limited amount of memory.

\subsection{\EFD Algorithm}\label{ssec:alg-construct}
\EFD is a randomized algorithm designed to perform memory-efficient detection of
low-rate large flows; it is designed to scale to a large number of flows, as
encountered by an Internet core router. \EFD is designed to limit the damage
inflicted by low-rate large flows while using very limited memory. The intuition
behind \EFD is to monitor subsets of flows, recursively subdividing the subset
deemed most likely to contain a large flow. By dividing subsets in this way,
\EFD exponentially reduces memory requirements (it can monitor $m^d$ flows with
$O(m+d)$ memory).

The main challenges addressed by \EFD include
efficiently mapping flows into recursively divided groups,
choosing the correct subdivision to reduce detection delay and FNs,
and configuring \EFD to guarantee the absence of FPs.

\paragraph{Recursive subdivision.}
To operate with limited memory, \EFD recursively subdivides monitored
flows into $m$ groups, and subdivides only the one group most likely to contain
a large flow.

\begin{figure}[!th]
\centering
\subfigure[\small{Virtual Counter Tree \newline (Full $m$-branch Tree)}] {
  \label{fig:efd-tree}
  \includegraphics[width=0.3\columnwidth, trim= 20 10 0 0]{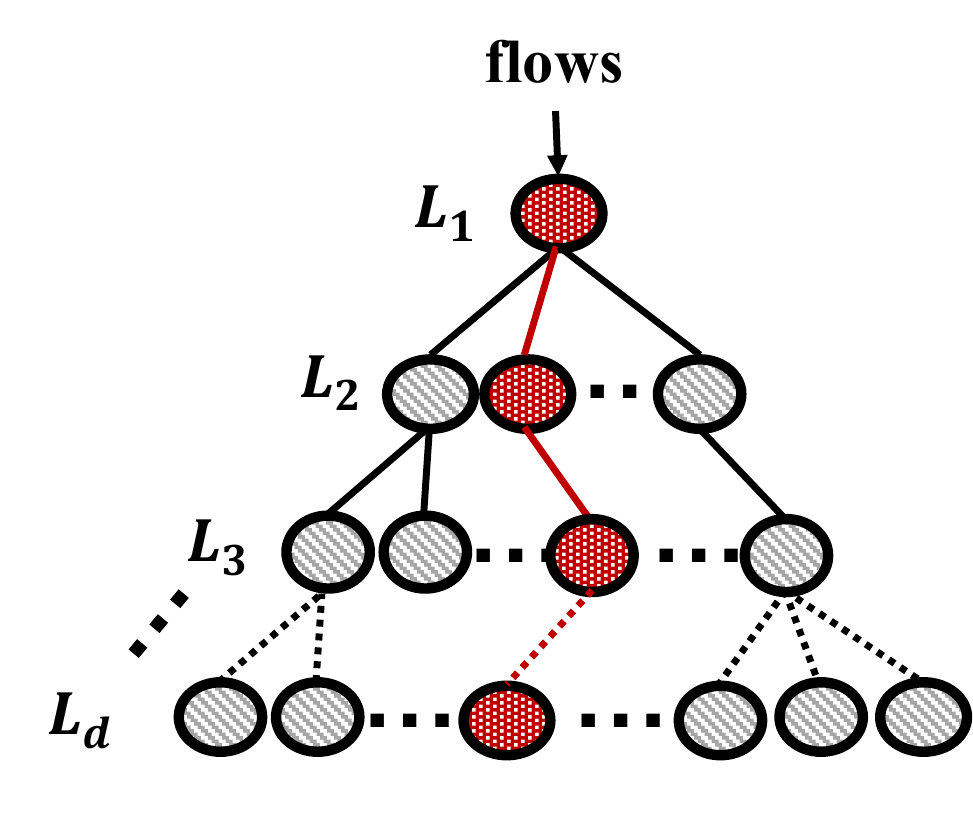}}
\subfigure[\small{A Tree \newline Branch.}] {
  \label{fig:efd-branch}
  \includegraphics[width=0.15\columnwidth, trim= 0 10 0 0]{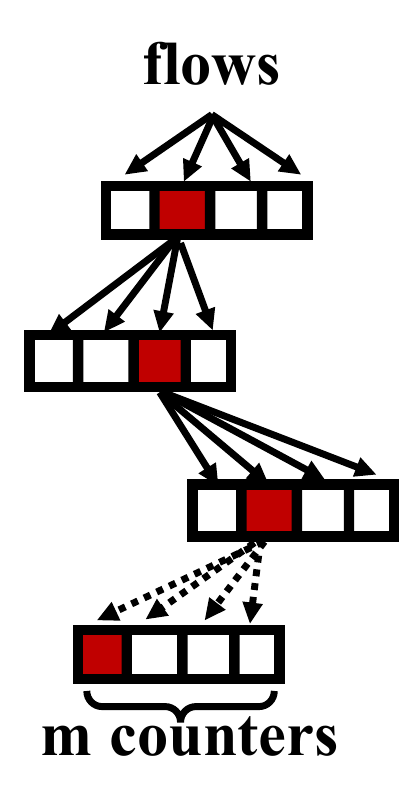}}
\subfigure[\small{Example with $7$ flows, \newline $m=4$, and $d=2$.}] {
  \label{fig:efd-example}
  \includegraphics[width=0.28\columnwidth, trim= 0 0 0 0]{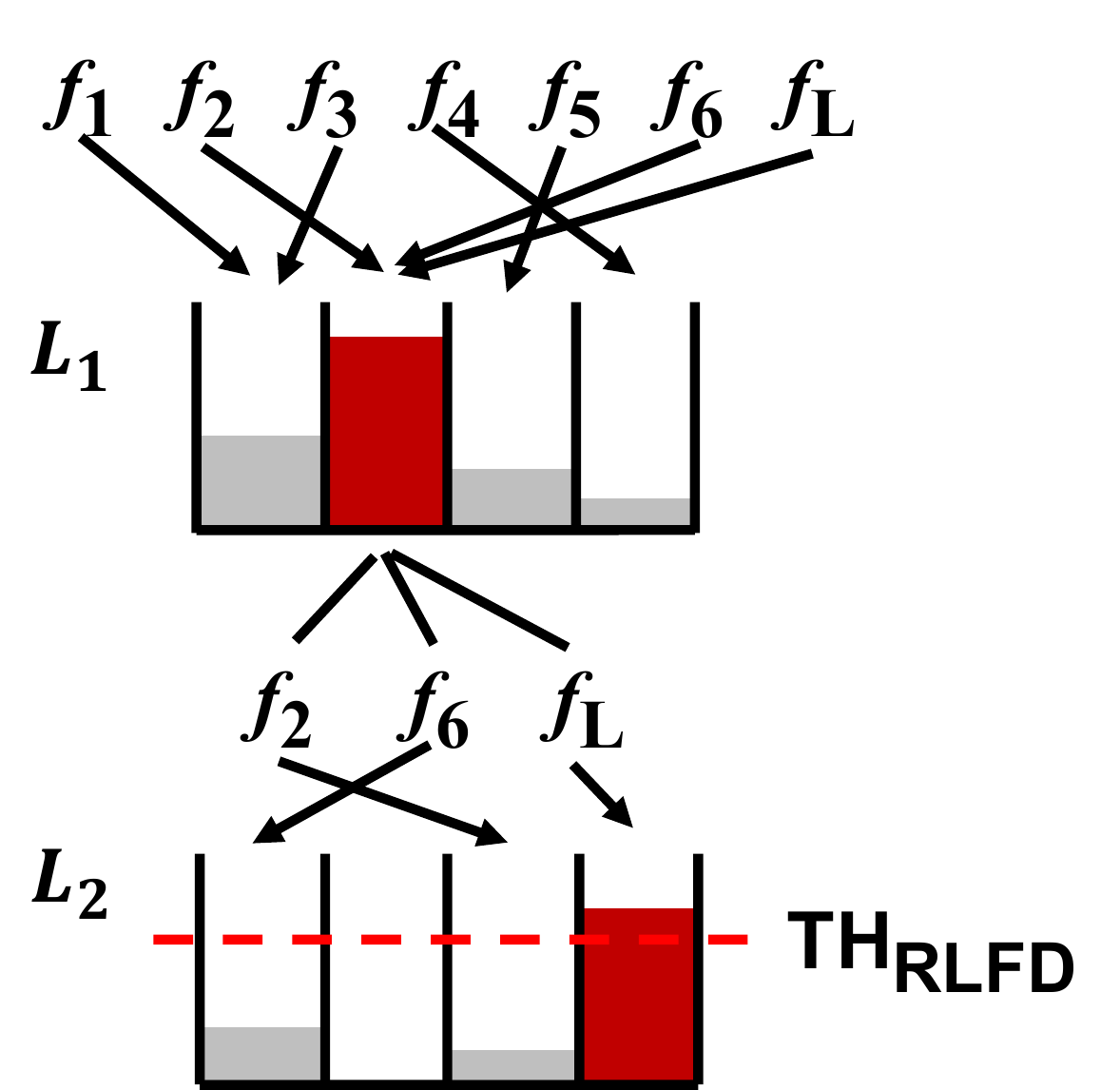}}
\vspace*{-4mm}
\caption{\EFD Structure and Example.}
\end{figure}

We can depict an \EFD as a \emph{virtual counter tree}\footnote{The terms
``counter tree'' and ``virtual counter'' are also used by
Chen~\etal~\cite{chen2016counter}, but our technique differs in both approach
and goal. Chen~\etal efficiently manage a sufficient number of counters for
per-flow accounting, while \EFD manages an insufficient number of counters to
detect consistent overuse.
}
(Figure~\ref{fig:efd-tree}) of depth $d$. Every non-leaf node in this tree has
$m$ children, each of which corresponds to a \emph{virtual counter}. The tree is
a full $m$-ary tree of depth $d$, though at any moment, only one node ($m$
counters) is kept in memory; the rest of the tree exists only virtually.

Each flow $f$ is randomly assigned to a path $\mit{PATH}(f)$ of counters on the
virtual tree, as illustrated by the highlighted counters
in Figure~\ref{fig:efd-branch}.
This mapping is determined by hashing a flow ID
with a keyed hash function,
where the key is randomly generated by each router.
\iftechreport
Section~\ref{ssec:efd-hash-impl}
\else
Our technical report~\cite{full-paper}
\fi
explains how \EFD efficiently implements this random mapping.

\vspace*{-7mm}
\vspace*{-5mm}

Since there are $d$ levels, each leaf node at level $L_d$ will contain an
average of $n/m^{d-1}$ flows, where $n$ is the total number of flows on the
link. A flow $f$ is identified as a large flow if it is the only flow associated
with its counter at level $L_d$ and the counter value exceeds a threshold
$\mit{TH}_{\fmit{\EFD}}$.
To reflect the flow specification $\func{TH}{t} = \gamma t + \beta$ from
Section~\ref{ssec:large-flow-detection}, we set $\mit{TH}_{\fmit{\EFD}} = \gamma
T_{\ell} + \beta$, where $T_{\ell}$ is the duration of the period during which
detection is performed at the bottom level $L_d$. Any flow sending more traffic
than $\mit{TH}_{\fmit{\EFD}}$ during any duration of time $T_{\ell}$ must
violate the large-flow threshold $\func{TH}{t}$, so \EFD has no FPs. We provide
more details about how we balance detection rate and the no-FP guarantee
\iftechreport
in Appendix~\ref{ssec:appendix-no-fp}.
\else
in our technical report~\cite{full-paper}.
\fi

\EFD considers only one node in the virtual counter tree at a time,
so it requires only $m$ counters.
To enable exploration of the entire tree,
\EFD divides the process into $d$ periods;
in period $k$, it loads one tree node from level $L_k$.
Though these periods need not be of equal length,
in this paper we consider periods of equal length $T_\ell$,
which results in a \EFD detection cycle $T_c = d\cdot T_{\ell}$.

\EFD always chooses the root node to monitor at level $L_1$;
after monitoring at level $L_k$,
\EFD identifies the largest counter
$C_{\fmit{max}}$ among the $m$ counters at level $L_k$,
and uses the node corresponding to that counter for level $L_{k+1}$.
\iftechreport
Section~\ref{ssec:efd-detection-prob} shows that
\else
Our technical report~\cite{full-paper} shows that
\fi
choosing the largest counter detects large flows with high probability.

Figure~\ref{fig:efd-example} shows an example with $m = 4$ counters,
$n = 7$ flows, and $d=2$ levels. $f_L$ is a low-rate large flow.
In level $L_1$, the largest counter is the one associated with
large flow $f_L$ and legitimate flows $f_2$ and $f_6$.
At level $L_2$, the flow set $\{f_L, f_2, f_6\}$
is selected and sub-divided.
After the second round, $f_L$ is detected
because it violates the counter value threshold $\mit{TH}_{\fmit{\EFD}}$.

\begin{figure}[!th]
\centering
\subfigure[\small{Level Change Diagram.}] {
  \label{fig:efd-level-diagram}
  \includegraphics[width=0.8\columnwidth, trim= 0 10 0 0]{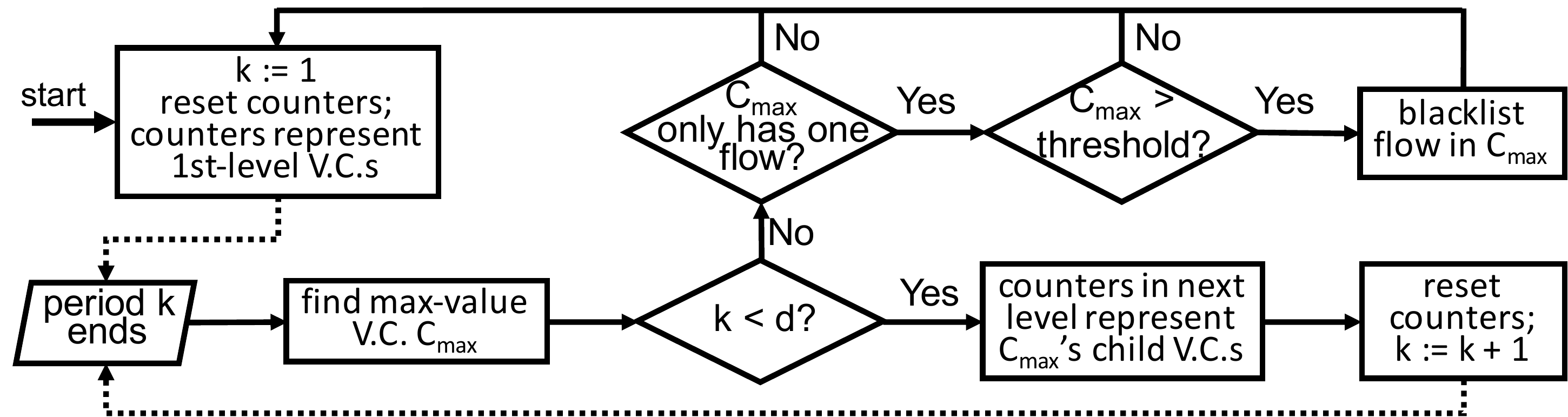}}
\subfigure[\small{Packet Processing Diagram.}] {
  \label{fig:efd-packet-diagram}
  \includegraphics[width=0.8\columnwidth, trim= 0 10 0 0]{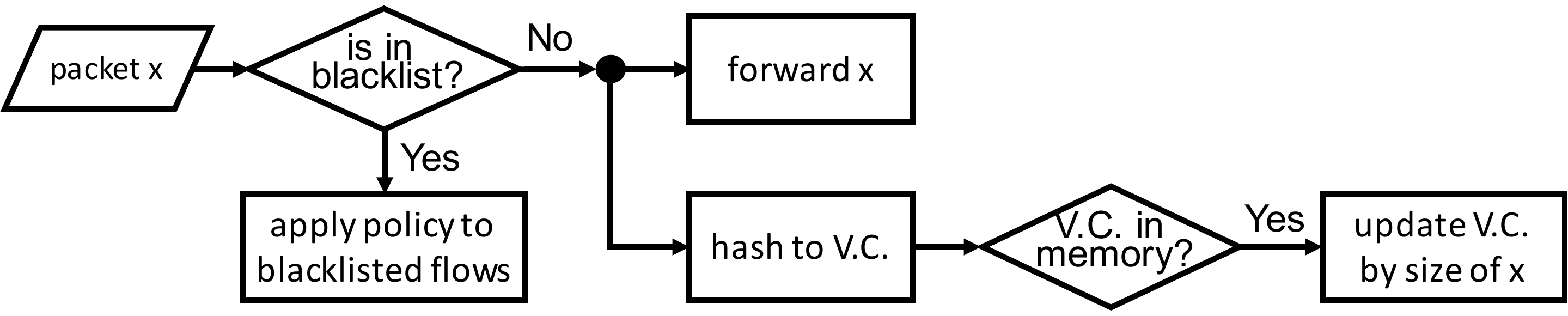}}
\vspace*{-4mm}
\caption{\EFD Decision Diagrams. ``V.C.'' stands for virtual counter.}\label{fig:efd-diagram}
\end{figure}

\paragraph{Algorithm description.}
As shown in Figure~\ref{fig:efd-level-diagram},
the algorithm starts at the top level $L_1$
so each counter represents a child of the root node.
At the beginning of each period, all counters are reset to zero.
At the end of each period,
the algorithm finds the counter holding the maximum value
and moves to the corresponding node,
so each counter in the next period is a child of that node.
Once the algorithm has processed level $d$,
it repeats from the first level.

Figure~\ref{fig:efd-packet-diagram} describes how \EFD processes each incoming
packet. When \EFD receives a packet $x$ from flow $f$, $x$ is dropped if $f$ is
in the \emph{blacklist} (a table that stores previously-found large flows). If
$f$ is not in the blacklist, \EFD hashes $f$ to the corresponding counters in
the virtual counter tree (one counter per level of the tree). If one such
counter is currently loaded in memory, its value is increased by the size of the
packet $x$. At the bottom level $L_d$, a large flow is identified when there is
only one flow in the counter and the counter value exceeds the threshold
$\mit{TH}_{\fmit{\EFD}}$. To increase the probability that a large flow is in a
counter by itself, we choose $d \ge \lceil \log_m n \rceil$ and use Cuckoo
hashing~\cite{pagh2001cuckoo} at the bottom level to reduce collisions.
Once a large flow is identified, it is blacklisted: in our evaluation we
calculate the damage $\damage$ with the assumption that large flows are blocked
immediately after having been added to the blacklist.

\vspace*{-6mm}
\vspace*{-6mm}

\iftechreport
\subsection{Implementation of Hashing and Counter Checking}
\label{ssec:efd-hash-impl}

Hashing each flow $f$ into a path of virtual counters $\mit{PATH}(f)$
and checking whether any of these counters are loaded in the
memory 
are two performance-critical operations of \EFD.

For each packet, our implementation only requires three bitwise operations (a
hash operation, a bitwise AND operation, and a comparison over $64$ bits), thus
requiring only $O(1)$ time\footnote{This is not entirely exact, as the length of
the hash output has to increase as $O(d\log m)$. However, in any realistic
scenario $d\log m$ is small enough to be considered constant.} and $O(1)$ space
on a modern $64$-bit CPU.

A naive implementation of hashing could introduce unnecessary cost in
computation and space. For example, a naive implementation may maintain one hash
function per virtual counter array. To check whether an incoming flow needs to
be monitored, it would have to check whether the incoming flow is hashed into
every maximum-value counter in each level above the current level. However, this
would take $O(d)$ time for checking level by level and $O(d)$ space for hash
functions, where $d$ is the depth of the virtual counter tree.

\begin{figure}[!th]
\centering
\includegraphics[width=0.8\columnwidth, trim= 0 20 0 0]{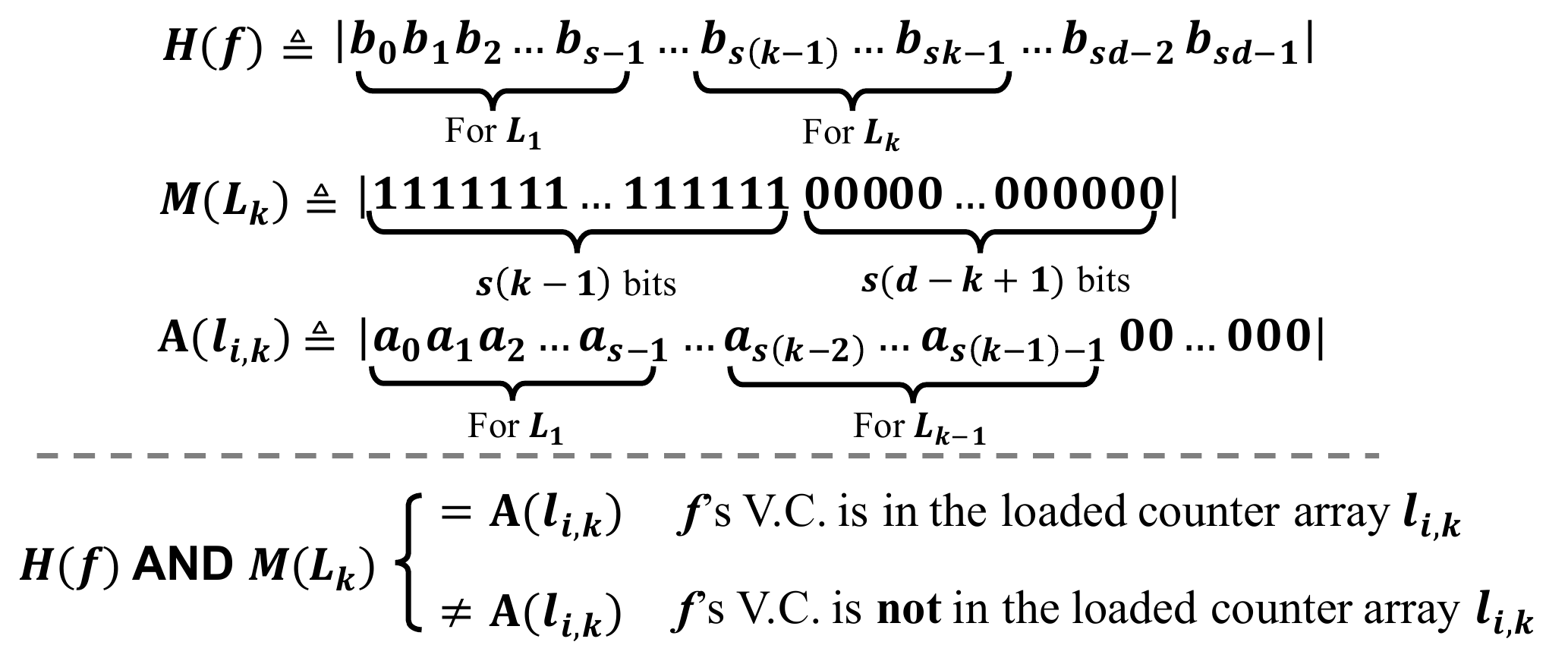}
\caption{\EFD Counter Hash and In-memory check.
$H(f)$ reflects the hash-generated bin number for all levels,
$M(L_k)$ reflects a mask that includes the first $k-1$ levels,
and $A(l_{i,k})$ reflects the bins selected in each of the first $k-1$ levels.
Flow $f$ is in the $i,k$ counter array exactly when
$H(f) \& M(L_k) = A(l_{i,k})$.}
\label{fig:efd-hash}
\end{figure}

Inspired by how a network router finds the subnet of an IP address, as
Figure~\ref{fig:efd-hash} illustrates, we map a flow to a virtual counter per
level based on a single hash value. Specifically, given an incoming flow $f$, we
compute $H(f)$, and then do a bitwise AND operation of $H(f)$ and a mask value
$M(L_k)$ of the current level $L_k$. We then check whether the result is equal to
the hash value $A(\ell_{i,k})$ of the currently loaded counter array $\ell_{i,k}$
(the $i$th counter array in the $k$th level). If the $H(f)$ AND
$M(L_k) = A(\ell_{i,k})$, then the virtual counter of $f$ in the
level $L_k$ is in the currently loaded counter array $\ell_{i,k}$.

Assuming \EFD has $d$ levels and $m$ counters in each counter array, we hash a
flow ID $f = \func{fid}{x}$ into $H(f)$ with $s\cdot d$ bits, where $s = \log_2
m$. We require the system designer to only choose the base-2 exponential value
for $m$, so that the $s$ is an integer.

The bits $[b_{s(k-1)}:b_{sk-1}]$
\footnote{$[b_i:b_j]$ denotes a block of bits $\{b_k\}$, $i \le k \le j$.}
of $H(f)$ are the index of the virtual counter in its counter array in the $k$th
level $L_k$.  As each counter array is determined by its ancestor counters as
Figure~\ref{fig:efd-branch} describes, the bits $[b_0:b_{s(k-1)-1}]$ can
uniquely determine the counter array in the level $L_k$ for the flow $f$.  Thus,
to check whether the virtual counters of a flow is in memory, we just need to
track the ancestor counters of the currently loaded counter array $\ell_{i,k}$.
We track the ancestor counters by $A(\ell_{i,k})$, which is also a value of
$s\cdot d$ bits. The bits $[a_0:a_{s(k-1)-1}]$ record the index of ancestor
counters of $\ell_{i,k}$, and the rest of bits are all $0$s. To track
$A(\ell_{i,k})$, we just simply set the bits $[a_{s(k-1)}:a_{sk-1}]$ as the
index of the selected counter at the end of the period of $L_k$.  The mask value
for the level $L_k$ is also a value of $s\cdot d$ bits, whose first $s(k-1)$
bits are $1$s and the rest are $0$s.  By $H(\func{fid}{x})$ AND $M(L_k)$, we
extract the ancestor bits $[b_0:b_{s(k-1)-1}]$ of the flow $\func{fid}{x}$, and
compare it with the ancestor bits $[a_0:a_{s(k-1)-1}]$ of the loaded counter
array.  If they match, then the flow $\func{fid}{x}$'s counter is in the memory,
and we update the counter with index $[b_{s(k-1)}:b_{sk-1}]$ by the size of the
packets of the flow $\func{fid}{x}$.

For each packet, our implementation above only need three basic operations: a
hash operation, an AND operation, and a comparison over $d\log_2 m$ bits.
Although the number of bits used in this implementation depends on $d$ and $m$,
a $64$-bit long integer is enough in most of the cases, thus those operations
only take O(1) CPU cycles in a modern $64$-bit CPU.

\fi

\subsection{\EFD Details and Optimization}
\label{ssec:efd-optimizations}
We describe some of the details of \EFD and propose additional optimizations to
the basic \EFD described in Section~\ref{ssec:alg-construct}.

\paragraph{Hash function update.}
We update the keyed hash function by choosing a new key at the beginning of
every initial level to guarantee that the assignment of flows to counters
between different top-to-bottom detection cycles is independent and
pseudo-random. For simplicity, in this paper we analyze \EFD assuming the random
oracle model. Picking a new key is computationally inexpensive and needs to be
performed only once per cycle.

\paragraph{Blacklist.}
When \EFD identifies a large flow,
 the flow's ID should be added to the blacklist as quickly as possible.
Thus, we implement the blacklist with a small amount of
L1 cache backed by permanent storage, e.g., main memory.
Because the blacklist write only happens at the bottom-level period
and the number of large flows detected in one iteration of the algorithm
is at most one, we first write these large flows in the L1 cache
and move them from L1 cache to permanent storage at a slower rate.
By managing the blacklist in this way,
we provide high bandwidth for blacklist writing,
defending against attacks that overflow the blacklist.

\paragraph{Using multiple \EFD{}s.}
If a link handles too much traffic to use a single \EFD,
we can use multiple \EFD{}s in parallel.
Each flow is hashed to a specific \EFD
so that the load on each detector meets performance requirements.
The memory requirements scale linearly in the number of \EFD{}s
required to process the traffic.

\iftechreport
\subsection{\EFD Runtime Analysis}
\label{ssec:efd-complexity}

We analyze the runtime using the same CPU considered in EARDet~\cite{eardet}.
An OC-768 ($40$~Gbps) high-speed link can accommodate $40$~million
mid-size ($1000$~bit) packets per second.
To operate at the line rate, a modern $3.2$~GHz CPU must process each packet within $76$~CPU cycles.
A modern CPU might contain
$32$~KB L1 cache,
$256$~KB L2 cache, and $20$~MB L3 cache.
It takes $4$, $12$, and $30$~CPU cycles to access L1, L2, and L3 CPU cache,
respectively; accessing main memory is as slow as $300$~cycles.

If, over a $40$-Gbps link,
we conservatively pick a large-flow threshold rate $\gamma=100$~kbps, a maximum
of $400,000$ flows can be supported. An \EFD with $400,000$ flows and
$128$ counters per level only needs $d=3$ levels
to get an average of $24.4$ flows at the bottom level, causing only a few
collisions for the $128$ counters at the bottom level which will be handled by
the Cuckoo hashing approach.
Even if we consider a much larger number of flows,
such as $40$~million,
$d=4$ levels results in around $19.1$ flows at the bottom level.
In such a $4$-level \EFD,
a flow's path through the tree will require
only $4\cdot \log_2 128 = 28$ bits,
so a $64$-bit integer is large enough for the hash value.
In practice, the threshold rate is higher than $100$~kbps,
and the number of flows is likely to be under $40$~million.

\paragraph{Computational complexity.}
Based on the implementation and optimizations
\iftechreport
in Sections~\ref{ssec:efd-hash-impl} and~\ref{ssec:efd-optimizations},
\else
in Section~\ref{ssec:efd-optimizations} and Appendix~\ref{ssec:efd-hash-impl},
\fi
\EFD performs the following steps on each packet:
\emph{(1) a hash computation} to find the flow's path in the tree,
\emph{(2) a bitwise AND operation}
  to find the subpath down to the depth of the current period,
\emph{(3) an integer comparison}
  to determine if the flow is part of an active counter,
and \emph{(4) a counter value update}
  if the flow is hashed into the loaded counter array.
Each of these operations is $O(1)$ complexity
and fast enough to compute within $76$ CPU cycles.

At the bottom level,
after operations \emph{(1)} to \emph{(3)},
\EFD performs the following steps:
\emph{(5) a Cuckoo lookup/insert} to find the appropriate counter,
\emph{(6) a counter value update} to represent the usage of a flow,
\emph{(7) a large-flow check} that compares the counter value with a threshold,
and \emph{(8) an on-chip blacklist write}
  if the counter has exceeded the threshold.
Steps \emph{(5)}--\emph{(7)} are only performed on packets
from the small fraction of flows that are loaded in the bottom-level array;
step \emph{(8)} is only for packets of the flows identified as large flows
in step \emph{(7)}, and this only happens once
for each flow (if we block the large flows in the blacklist).
Thus steps \emph{(5)}--\emph{(8)} are executed much less frequently
than steps \emph{(1)}--\emph{(4)}.
Even so, steps \emph{(5)}--\emph{(8)} have a constant time in expectation,
and are negligible in comparison with steps \emph{(1)}--\emph{(4)}.

\paragraph{Storage complexity.}
\EFD only keeps a small array of counters
and a few additional variables:
the hash function key, the $64$-bit mask value for the current level,
and the $64$-bit identifier of the currently loaded counter array.
Because we use Cuckoo hashing at the bottom level, besides
a $32$-bit field for the counter value,
each counter entry needs to have a field for the associated flow ID key,
which is $96$ bits in IPv4 and $288$ bits in IPv6.
An array of $128$~counters requires $2$~KB in IPv4 and $5$~KB for IPv6,
which readily fits within the L1 cache.
\iftechreport
As discussed in Appendix~\ref{ssec:appendix-shrink-counter-size},
we can further shrink
\else
Our technical report describes an optimization that shrinks
\fi
the flow ID field size to $48$ bits
(with FP probability $\le 2^{-38}$ for each flow);
if deployed, a $128$~counter array is $1.25$~KB
and a $1024$~counter array is $10$~KB for both IPv4 and IPv6,
which can fit into the L1 cache ($32$ KB).

\fi

\subsection{\EFD's Advantages and Disadvantages}
\paragraph{Advantages.}
With recursive subdivision and additional optimization techniques,
\EFD is able to (1) identify low-rate large flows with non-zero probability,
with probability close to 100\% for flows that cause extensive damage
\iftechreport
(Section~\ref{ssec:efd-detection-prob}
\else
(our technical report~\cite{full-paper}
\fi
analyzes \EFD's detection probability); and (2) guarantee no-FP, eliminating
damage due to FP.

\paragraph{Disadvantages.}
First, a landmark-window-based algorithm such as \EFD cannot guarantee exact
detection over large-flow specification based on arbitrary time
windows~\cite{eardet} (landmark window and arbitrary window are
introduced in Section~\ref{ssec:large-flow-detection}).
However, this approximation results in limited damage,
as mentioned in Section~\ref{sec:overview}.
Second, recursive subdivision based on landmark time windows
requires at least one detection cycle to catch a large flow.
Thus, \EFD cannot guarantee low damage for flows with very high rates.
Third, \EFD works most effectively when the large flow
exceeds the flow specification in all $d$ levels,
so bursty flows with a burst duration
shorter than the \EFD detection cycle $T_c$ are likely to escape detection
(where \emph{burst duration} refers to the amount of time during which
the bursty flow sends in excess of the flow specification).

\subsection{\name Hybrid Scheme}\label{ssec:eardet-efd-hybrid}
We propose a hybrid scheme, \name,
which is a parallel composition with one EARDet and two \EFDs (\twinEFD).
This hybrid can detect both high-rate and
low-rate large flows without producing FPs,
requiring only a limited amount of memory.
We use EARDet instead of Flow Memory in this hybrid scheme because
EARDet's detection is deterministic, thus has shorter detection delay.

\paragraph{Parallel composition of EARDet and \EFD.}
As described in Section~\ref{ssec:why-hybrid},
we combine EARDet and \EFD in parallel
so that \EFD can help EARDet detect low-rate flat flows,
and EARDet can help \EFD quickly catch
high-rate flat and bursty flows.

\paragraph{\twinEFD parallel composition.}
\EFD is most effective at catching flows
that violate flow specification across an entire detection cycle $T_c$.
An attacker can reduce the probability of being caught by \EFD
by choosing a burst duration shorter than $T_c$
and an inter-burst duration greater than $T_c/d$
(thus reducing the probability that the attacker
will advance to the next round during its inter-burst period).
We therefore introduce a second \EFD
(\EFD{}$^{(2)}$) with a longer detection cycle
$T_c^{(2)}$, so that a flow must have
burst duration shorter than $T_c^{(1)}$
and burst period longer than $T_c^{(2)}/d$
to avoid detection by the \twinEFD
(where \EFD{}$^{(1)}$ and $T_c^{(1)}$,
are the first \EFD and its detection cycle respectively).
For a given average rate,
flows that evade \twinEFD have a higher burst rate
than flows that evade a single \EFD.
By properly setting $T_c^{(1)}$ and $T_c^{(2)}$,
\twinEFD can synergize with EARDet,
ensuring that a flow undetectable by \twinEFD
must use a burst higher than
EARDet's rate threshold $\gamma_h$.

\paragraph{Timing randomization.}
An attacker can strategically send traffic
with burst durations shorter than $T_c^{(1)}$,
but choose low duty cycles to avoid detection by both
\EFD{}$^{(1)}$ and EARDet.
Such an attacker can only be detected by \EFD{}$^{(2)}$,
but \EFD{}$^{(2)}$ has a longer detection delay,
allowing the attacker to maximize damage before being blacklisted.
To prevent attackers from deterministically maximizing damage,
we randomize the length of the detection cycles $T_c^{(1)}$ and $T_c^{(2)}$.

\iftechreport
\section{Theoretical Analysis}
\label{sec:analysis}
\iftechreport  
\begin{table}[htbp]\caption{Table of Notations.}\label{table:notation}
\begin{small}
    \begin{tabular}{r c p{6cm} }
      \toprule
      \multicolumn{3}{l}{\emph{Generic notations:}}\\
      $\rho$ & $\triangleq$ & Rate of (outbound) link\\
      $\gamma, \beta$ & $\triangleq$ & Rate and burst threshold flow specification\\
      $\theta$ & $\triangleq$ & Duty cycle of bursty flows ($\theta \le 1$)\\
      $T_b$ & $\triangleq$ & Period of burst\\
      $R_{\fmit{atk}}, \alpha$ & $\triangleq$ & Average large-flow rate, and $R_{\fmit{atk}} = \alpha\gamma$\\
      $n$ & $\triangleq$ & Number of legitimate flows\\
      $n_{\gamma}$ & $\triangleq$ & $\frac{\rho}{\gamma}$; Maximum number of legitimate flows at rate $\gamma$\\
      $m$ & $\triangleq$ & Number of counters available in a detector\\
      $\gamma_{h}$ & $\triangleq$ & $\frac{\rho}{m+1}$; EARDet high-rate threshold rate\\
      $E(D_{\fmit{over}})$ & $\triangleq$ & Expected overuse damage\\
      \hline 

      \multicolumn{3}{l}{\emph{\EFD notations:}}\\
      $d$ & $\triangleq$ & Number of levels\\
      $n^{(k)}$ & $\triangleq$ & Number of legitimate flows in the level $k$\\
      $T_{\ell}$ & $\triangleq$ & Time period of a detection level\\
      $T_c$ & $\triangleq$ & Detection cycle $T_c = d\cdot T_{\ell}$\\
      $Pr(A_{\alpha})$ & $\triangleq$ & Detection prob. for flows with $R_{\fmit{atk}}=\alpha\gamma$\\      
      $\alpha_{0.5}$ & $\triangleq$ & When $\alpha\ge\alpha_{0.5}$, approximately $Pr(A_{\alpha}) \ge 0.5$\\
      $\alpha_{1.0}$ & $\triangleq$ & When $\alpha\ge\alpha_{1.0}$, approximately $Pr(A_{\alpha}) = 1.0$\\

      \bottomrule
    \end{tabular}
\end{small}
\end{table}
\fi 

\iftechreport
In this section, we
\else
We now
\fi
discuss \EFD's performance and its large-flow detection probability.
We then compare \name with state-of-the-art schemes, considering various types
of large flows under \name's worst-case background traffic. Due to limited
space, some derivations are in
\iftechreport
the appendix.
\else
our technical report~\cite{full-paper}.
\fi

\paragraph{Detection probability.}
\emph{Single-level detection probability} is the probability that
a \EFD selects a correct counter (containing at least one large flow)
for the next level.
\emph{Total detection probability} is the probability 
that one copy of \EFD catches a large flow in a cycle $T_c$,
which is the product of the single-level detection probabilities
across all levels in a cycle, minus the probability that two large flows will
be assigned to the highest counter at the last level. The subtrahend is small
and negligible when the number of levels is large enough.

\subsection{\EFD Worst-case Background Traffic}\label{ssec:efd-ball-into-bins}

Since our goal is to minimize worst-case damage, we assume the
worst-case background traffic against \EFD in the rest of the
analysis. Given a large flow, the worst-case background traffic is the
legitimate flow traffic pattern that maximizes damage caused by a
large flow. Since damage increases with expected detection delay (and thus
decreases with single-level detection probability) in \EFD, we
derive the worst-case background traffic by finding the minimum
single-level detection probability for each level of \EFD.
Theorem~\ref{th:efd-worst-case} states that the worst-case
background traffic consists of threshold-rate legitimate flows fully
utilizing the outbound link. The proof and discussion of
Theorem~\ref{th:efd-worst-case} and are presented
\iftechreport
    in Appendix~\ref{ssec:appendix-worst-efd}.
\else
    in our technical report~\cite{full-paper}.
\fi

\begin{theorem}\label{th:efd-worst-case}
On a link with a threshold rate $\gamma$ and an outbound link capacity
$\rho$, given an attack large flow $f_{\fmit{atk}}$, \EFD runs with the lowest
probability to select the counter containing $f_{\fmit{atk}}$ to the next level, when
there are $n_{\gamma}=\rho/\gamma$ legitimate flows, each of which is at the  rate of $\gamma$.
\end{theorem} 

\iftechreport
Figure~\ref{fig:worst-background} in Appendix~\ref{ssec:appendix-worst-efd}
presents single-level detection probabilities for several different background
traffic patterns, which empirically validates our theorem.
\fi

\subsection{Characterizing Large Flows}
To systematically compare \name with other detectors under various
types of attack flows, we categorize large flows based on three
characteristics, as Figure~\ref{fig:flow-def} illustrates:
\begin{enumerate}
\item \emph{Burst Period} ($T_b$). A large flow sends 
a burst of traffic in a period of $T_b$.
\item \emph{Duty Cycle} ($\theta \in (0,1]$). 
In each period of length $T_b$, a large flow only sends packets during a continuous 
time period  of $\theta T_b$ and remains silent during the rest of the period.
\item \emph{Average Rate} ($R_{\fmit{atk}}$). This is the average volume of 
traffic sent from a large flow per second over a time interval much longer than
the burst period $T_b$. The instant rate during the burst chunk $\theta T_b$
is $R_{\fmit{atk}}/\theta$.
\end{enumerate}

\begin{figure}[!th]
\centering
\includegraphics[width=0.5\columnwidth]{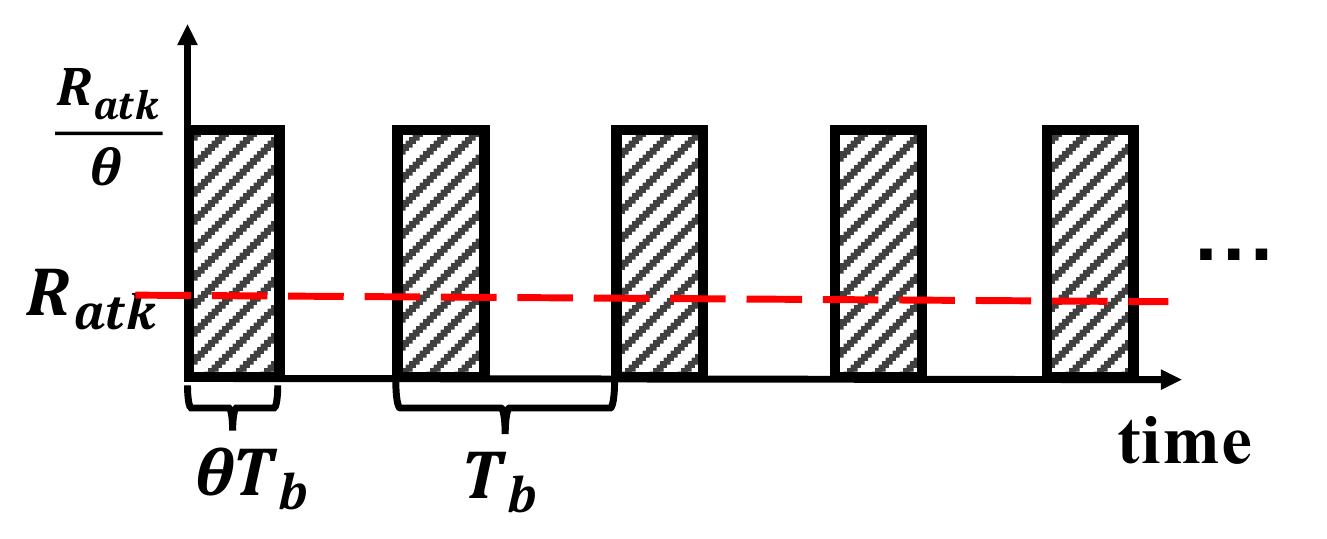}
\vspace*{-3mm}
\caption{Flow with average rate $R_{\fmit{atk}}$, burst period $T_b$,
duty cycle $\theta \in (0, 1]$.
}
\label{fig:flow-def}
\end{figure}

By remaining silent between bursts, attacks such as 
the Shrew attack~\cite{Kuzmanovic2003} keep the average rate lower than 
the detection threshold to evade the detection algorithms based on 
landmark windows~\cite{Misra1982, Demaine2002, Karp2003a, Manku2002, 
metwally2005efficient, Estan2003, Fang1999, Cormode2005}.

A large flow may switch between different characteristic patterns over
time, including ones that comply with  flow specifications. The
total damage in this case can be computed by adding up the damage
inflicted by the large flow under each appearing pattern. Hence, for
the purpose of the analysis, we focus our discussion on large flows
with fixed characteristic patterns.

\iftechreport  
\subsection{\EFD Detection Probability for Flat Flows}
\label{ssec:efd-detection-prob}

In order to detect a flat ($\theta = 1$) large flow, the traffic of the flat
large flow should be observable in each detection level.

The probability that \EFD catches one large flow in a detection cycle increases
with the number of large flows passing through \EFD.
Because a greater number of large flows implies that more counters may contain
large flows in each level, \EFD has a higher chance of correctly selecting
counters with large flows in the recursive subdivision. 
We therefore discuss the worst-case scenario for \EFD 
where only one large flow is present.

Because the operation in all but the bottom level of \EFD is similar and the only difference is the flows hashed to the counter array, we
discuss the detection in a single level first and expand it to the whole
detection cycle. Additional numeric examples are 
provided in Appendix~\ref{ssec:appendix-efd-detection-prob}.

\paragraph{Single-level detection probability.}
Given the total number of flows traversing the link is $n$, we can 
predict the expected number of flows in the $k$th level by
$n^{(k)} = n/m^{k-1}$, where $m$ is the number of counters.
Since $n^{(k)}$  depends only on the total number of flows
and not the traffic distribution,
we discuss a single-level detection with $n^{(k)}$ legitimate flows,
$m$ counters, and a large flow at the rate of $R_{\fmit{atk}}=\alpha\gamma$,
where $\gamma$ is the threshold rate and $\alpha > 1$.
When the context is clear, we use $n$ to stand for $n^{(k)}$ in the
discussion of single-level detection.

According to Theorem~\ref{th:efd-worst-case}, 
the worst-case background traffic is that
all $n$ legitimate flows are at the threshold rate $\gamma$; Theorem~\ref{th:efd-level-prob} shows 
an approximate lower bound of the single-level detection probability
$\func{P_{\fmit{worst}}}{m, n, \alpha}$ in such worst-case background traffic.
The proof of Theorem~\ref{th:efd-level-prob} and its 
Corollaries~\ref{cl:0.5-lower-bound} and~\ref{cl:1.0-lower-bound} are presented
in Appendix~\ref{ssec:appendix-efd-single-level-prob-proof}.

\begin{theorem}\label{th:efd-level-prob}
Given $m$ counters in a level, $n$ legitimate flows at full rate $\gamma$, 
and a large flow $f_{\fmit{atk}}$ with an average rate of 
$R_{\fmit{atk}}=\alpha\gamma$,
the probability $\func{P_{\fmit{worst}}}{m, n, \alpha}$ that 
 \EFD will correctly select the counter with large flow $f_{\fmit{atk}}$
has an approximate lower bound of $1 - Q(K, \frac{n}{m})$, 
where $K = \bigl\lfloor \frac{n}{m} + \sqrt{2\frac{n}{m}\log n} - \alpha \bigr\rfloor$;
$Q(K, \frac{n}{m})$ is the cumulative distribution function (CDF) 
of the Poisson distribution $\mit{Pois}(\frac{n}{m})$.
\end{theorem}

\begin{corollary}\label{cl:0.5-lower-bound}
For a detection level with $n$ legitimate flows, $m$ counters, and 
a large flow $f_{\fmit{atk}}$ at the average rate of 
$\alpha_{0.5} \cdot \gamma$, the probability 
$\func{P_{\fmit{worst}}}{m, n, \alpha_{0.5}}$ that \EFD will correctly select 
the counter of $f_{\fmit{atk}}$ has an approximate lower bound of $0.5$,
where $\alpha_{0.5} = \sqrt{2\frac{n}{m}\log n}$.
\end{corollary}

\begin{corollary}\label{cl:1.0-lower-bound}
For a detection level with $n$ legitimate flows, $m$ counters, and 
a large flow $f_{\fmit{atk}}$ at the average rate of 
$\alpha_{1.0} \cdot \gamma$, the probability 
$\func{P_{\fmit{worst}}}{m, n, \alpha_{1.0}}$ that \EFD will correctly select 
the counter of  $f_{\fmit{atk}}$ has an approximate lower bound of $1.0$,
where $\alpha_{1.0} = 2\cdot \alpha_{0.5} = 2\sqrt{2\frac{n}{m}\log n}$.
\end{corollary}

\paragraph{Total detection probability.}
Theorem~\ref{th:total-lower-bound} describes the total probability of detecting a
large flow in one detection cycle. Detailed proof is provided in
Appendix~\ref{ssec:appendix-efd-total-prob-proof}.

\begin{theorem}\label{th:total-lower-bound}
When there are $n$ legitimate flows and a flat large flow at the rate of 
$\alpha\gamma$, the total detection probability of a \EFD with
$m$ counters has an approximate lower bound:
\begin{small}
\begin{equation}\label{eq:approx-lower-bound-total-prob}
Pr(A_{\alpha}) \ge \left\{ \begin{aligned}
&\bigg(1 - Q(K_{\gamma}, \frac{n_{\gamma}}{m})\bigg)
    ^{\lfloor \log_m (n/n_{\gamma}) \rfloor + 1}
    \mit{, when $n \ge n_{\gamma}$}\\
&1 - Q(K, \frac{n}{m}) \hspace{21 mm} \mit{, when $n < n_{\gamma}$}\\
\end{aligned} \right.
\end{equation}
\end{small}
\noindent
where $K_{\gamma} = \bigl\lfloor \frac{n_{\gamma}}{m} + \sqrt{2\frac{n_{\gamma}}{m}\log n_{\gamma}} - \alpha \bigr\rfloor$ 
, $K = \bigl\lfloor \frac{n}{m} + \sqrt{2\frac{n}{m}\log n} - \alpha \bigr\rfloor$,
and $Q(x,\lambda)$ is the CDF of the Poisson distribution $\mit{Pois}(\lambda)$.
\end{theorem}

\subsection{\twinEFD Theoretical Overuse Damage}
\label{ssec:efd-damage}
To evaluate \EFD's performance, we derive a theoretical bound on the damage
caused by large flows against \EFD. 
Recall that there are two sources of
damage: FP damage $D_{\fmit{fp}}$ and overuse damage $D_{\fmit{over}}$. 
Because \EFD has no FP, there is no need to consider $D_{\fmit{fp}}$. 
Thus, we only theoretically analyze $D_{\fmit{over}}$.

Theorem~\ref{th:twin-efd-bursty-damage} shows the expected overuse damage for
flat flows and bursty flows against a \twinEFD. The proof is presented in
Appendix~\ref{ssec:appendix-efd-damage-proof}. Additional numeric examples
are in Appendix~\ref{ssec:appendix-efd-damage}. 

\begin{theorem}\label{th:twin-efd-bursty-damage}
 A \twinEFD with $\EFD^{(1)}$ and $\EFD^{(2)}$ whose detection cycles are
$T_c^{(1)}$ and $T_c^{(2)}=\frac{2d\gamma_h}{\alpha\gamma}T_c^{(1)}$,
respectively, it can detect bursty flows at an average rate 
$R_{\fmit{atk}} = \alpha\gamma<\theta\gamma_h$,
where $\gamma_h$ is the high-rate threshold rate of the EARDet.  The
expected overuse damage caused by such flows has the following upper bound:
\vspace*{-1mm}
\begin{small}
\begin{equation}\label{eq:twin-efd-bursty-damage}
E(D_{\fmit{over}}) \le \left\{ \begin{aligned}
&T_c^{(1)} \gamma\alpha / \theta Pr(A_{\alpha})
 \hspace{15 mm} \mit{, when $\theta T_b \ge 2T_c^{(1)}$}\\
&T_c^{(1)} 2d\gamma_h / \theta Pr(A_{\alpha})
 \hspace{15 mm} \mit{, when $\theta T_b < 2T_c^{(1)}$}\\
\end{aligned} \right.
\end{equation}
\end{small}
\vspace*{-1mm}
\noindent
where
\vspace*{-3mm}
\begin{small}
\begin{equation}
Pr(A_{\alpha}) \ge \left\{ \begin{aligned}
&\bigg(1 - Q(K_{\gamma}, \frac{n_{\gamma}}{m})\bigg)
    ^{\lfloor \log_m (n/n_{\gamma}) \rfloor + 1}
    \mit{, when $n \ge n_{\gamma}$}\\
&1 - Q(K, \frac{n}{m}) \hspace{21 mm} \mit{, when $n < n_{\gamma}$}\\
\end{aligned} \right.
\end{equation}
\end{small}
\noindent
and $K_{\gamma} = \bigl\lfloor \frac{n_{\gamma}}{m} + \sqrt{2\frac{n_{\gamma}}{m}\log n_{\gamma}} - \alpha_{\theta} \bigr\rfloor$ 
, $K = \bigl\lfloor \frac{n}{m} + \sqrt{2\frac{n}{m}\log n} -  \alpha_{\theta} \bigr\rfloor$ ($\alpha_{\theta}=\alpha/\theta$ when $\theta T_b \ge 2T_c^{(1)}$,
and $\alpha_{\theta}=\alpha$ when $\theta T_b < 2T_c^{(1)}$).
The $d$ is the number of levels in \EFD,
and $Q(x,\lambda)$ is the CDF of the Poisson distribution $\mit{Pois}(\lambda)$.
The damage of flat flow is that in the case of $\theta = 1$ and
$\theta T_b \ge 2T_c^{(1)}$.
\end{theorem}

We can see that a properly configured \twinEFD can detect bursty flows 
unable to be detected by EARDet
(i.e., flows at average rate $R_{\fmit{atk}} = \alpha\gamma<\theta\gamma_h$).

\fi

\begin{table*}[htbp]
\caption{Theoretical Comparison. \ETE outperforms other detectors
with lower large flow damage. Damage in megabyte (MB).
} 
\label{table:th-compare}
\centering
\begin{small}
  \begin{tabular}{|c|c|c|c|c|c|c|c|c|c|}
    \hline
    \hline
    & \multirow{3}{*}{Algorithm} &
    \multirow{2}{*}{FP}
    & \multicolumn{7}{|c|}{Overuse Damage (MB)}\\
    \hhline{~~~-------}
    & & Damage
    & \multicolumn{5}{|c|}{Low-rate Large Flow}
    & \multicolumn{2}{|c|}{High-rate Large Flow}\\
    \hhline{~~~-------}
    & & & \multirow{2}{*}{$R_{\fmit{atk}} < 10\gamma$}
    & \multicolumn{2}{|c|}{$10\gamma < R_{\fmit{atk}} < 30\gamma$}
    & \multicolumn{2}{|c|}{$30\gamma \le R_{\fmit{atk}} < 250\gamma$}
    & \multicolumn{2}{|c|}{$250\gamma \le R_{\fmit{atk}}$} \\
    \hhline{~~~~------} 
    & & & & $\theta T_b < 2T_c$ & $\theta T_b \ge 2T_c$
    & $\theta T_b < 2T_c$ & $\theta T_b \ge 2T_c$ 
    & $\theta T_b < 2T_c$ & $\theta T_b \ge 2T_c$\\
    \hline
    \multirow{4}{*}{\scriptsize{Individual}} 
    & \twinEFD & 0 & $[512, +\infty)$ & $[158, 512)$ & $[33, 45)$ & $[70, 158)$
    & $[6, 33)$ & $[99, +\infty)$ & $[6, +\infty)$\\
    \cline{2-10}
    & EARDet & $0$ & $+\infty$ & $+\infty$ & $+\infty$ 
    & $+\infty$ & $+\infty$ & $\approx 0$ & $\approx 0$\\
    \cline{2-10}
    & FM & $0$ & $+\infty^{*}$ 
    & $+\infty^{*}$ & $+\infty^{*}$ 
    & $+\infty^{*}$ & $+\infty^{*}$ & $\approx 0$ & $\approx 0$\\
    \cline{2-10}
    & AMF & $+\infty$ & $\approx 0$ & $\approx 0$ 
    & $\approx 0$ & $\approx 0$ & $\approx 0$ & $\approx 0$ & $\approx 0$\\
    \hline
    \hline
    \multirow{2}{*}{\scriptsize{Hybrid}}
    & \ETE & 0 & $[512, +\infty)$ & $[158, 512)$ & $[33, 45)$ & $[70, 158)$ 
    & $[6, 33)$ & $\approx 0$ & $\approx 0$\\
    \cline{2-10}
    & AMF-FM & $0$ & $+\infty^{*}$ 
    & $+\infty^{*}$ & $+\infty^{*}$ 
    & $+\infty^{*}$ & $+\infty^{*}$ & $\approx 0$ & $\approx 0$\\
    \hline
  \end{tabular}

  Comparison in a $40$ Gbps link with threshold rate $\gamma = 400$ Kbps.
  Each of \twinEFD, EARDet, FM and AMF has $m=100$ counters 
  (each of single \EFD has $50$ counters), 
  and thus each of \ETE and AMF-FM
  has $200$ counters. In \twinEFD and \ETE, 
  detection cycles $T_c^{(1)}=T_c=0.1$ sec,
  $T_c^{(2)} = 7.92$ sec, and number of levels is $d=4$. 
  Attack Flows are busty flows with duty cycle of $\theta=0.25$.
\iftechreport
  The reasons for this \twinEFD configuration are shown
  in Appendix~\ref{ssec:appendix-efd-damage}.
\fi

  $^{*}$The overuse damage for FM is treated as infinity, 
  due to the extremely low detection probability. 

\end{small}
\end{table*}

\subsection{Theoretical Comparison}

We compare the \ETE hybrid scheme with the most relevant competitor,
the AMF-FM hybrid scheme~\cite{Estan2003}, which runs an AMF
and a FM sequentially: all traffic is first sent to the AMF and the AMF
sends detected large flows (including FPs) to the
FM to eliminate FPs.
For completeness, we also present the results of
individual detectors, including \twinEFD, EARDet, AMF, and Flow Memory
(FM). Table~\ref{table:th-compare} summarizes the damage inflicted by
different large-flow patterns when different detectors are deployed.
\iftechreport
The damage is calculated according to the analyses of
AMF (Appendix~\ref{ssec:appendix-mf-analysis}),
FM (Appendix~\ref{ssec:appendix-fm-analysis}), EARDet~\cite{eardet},
and \twinEFD (Section~\ref{ssec:efd-damage}).
Figures~\ref{fig:efd-td-theta-025} and~\ref{fig:efd-td-theta-025-twin-efd}
in Appendix~\ref{ssec:appendix-efd-damage} 
provide more details about \twinEFD's
overuse damage presented in Table~\ref{table:th-compare}.
\fi

\paragraph{Comparison setting.}
To compare detectors in an in-core router setting, we allocate only
$100$ counters for each detector, and we allocate $50$ counters for
each \EFD in the \twinEFD for a fair comparison. Each hybrid scheme
has $200$ counters in total to ensure fair comparison between hybrid schemes is fair. 

We consider both high-rate large flows ($R_{\fmit{atk}} \geq
250\gamma$) and low-rate large flows ($R_{\fmit{atk}} <
250\gamma$).
 $250\gamma$ is the minimum rate 
at which detection is guaranteed by EARDet, FM, and AMF-FM:
$\frac{\theta\rho}{m} = \frac{0.25\times10^5\gamma}{100}$.
 Low-rate large flows are further divided into three
rate intervals for thorough comparison. For each rate interval, we
consider the worst-case ($\theta T_b < 2T_c$) and non-worst-case
($\theta T_b \ge 2T_c$) burst length. The duty cycle of the bursty
flow is set to $\theta = 0.25$, which is challenging
for \name. Given an average rate $R_{\fmit{atk}}$, 
if $\theta$ is close to $0$ (close to $1$), 
a bursty flow is easily detected
by EARDet (\twinEFD) in \name.

\paragraph{\name ensures lower damage.}
As shown in Table~\ref{table:th-compare}, \twinEFD and \ETE outperform
other detectors for identifying a wide range of low-rate flows. However,
due to limited memory, it remains challenging for \twinEFD and \ETE to
effectively detect large flows that are extremely close to the threshold.

We can see that \twinEFD fails to limit the damage caused by high-rate
large flows, because the overuse damage is linear in
$R_{\fmit{atk}}$ of high-rate flows (due to the minimum detection
delay of one cycle). Thus, \ETE uses EARDet to limit the damage
caused by high-rate flows.
\ETE is better than the AMF-FM hybrid scheme. This is because the FP from AMF (with limited memory) is too high
to narrow down the traffic passed to the FM in the downstream, so that
the FM's performance is not improved.

\begin{figure}[!th]
\centering
\subfigure[\small{When $\rho = 10^5\gamma$}] {
  \label{fig:detectable-rate-1}
  \includegraphics[width=0.45\columnwidth, trim= 0 0 0 0]{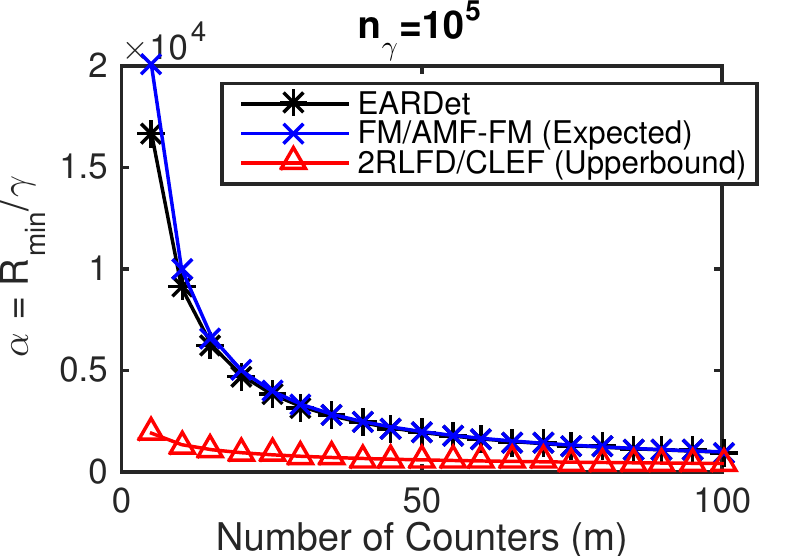}}
\subfigure[\small{When $\rho = 10^7\gamma$}] {
  \label{fig:detectable-rate-2}
  \includegraphics[width=0.45\columnwidth, trim= 0 0 0 0]{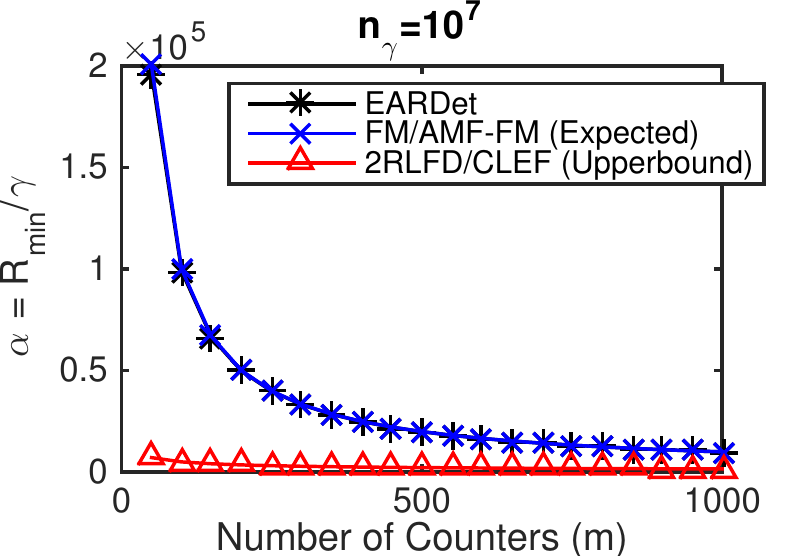}}
\vspace*{-4mm}
\caption{Minimum Rate of Guaranteed Detection $R_{\fmit{min}}$ 
(shown as $R_{\fmit{min}}/\gamma$ in figures),
for flat large flows ($\theta = 1.0$), 
when link capacity $\rho=10^5\gamma$ and $10^7\gamma$,
where $\gamma$ is threshold rate. 
\twinEFD and \ETE have much lower rate of guaranteed detection 
than other schemes when the memory is limited.}
\end{figure}

\paragraph{\name is memory-efficient.}
We now consider the minimum rate of guaranteed detection ($R_{\fmit{min}}$) for 
flat flows (i.e., flat large flows ($\theta=1.0$)
exceeding the rate $R_{\fmit{min}}$) of these
detectors. The $R_{\fmit{min}}$ of 
 \twinEFD and \ETE is bounded from above by
$4\theta \sqrt{\frac{m\log n_{\gamma}}{n_{\gamma}}}\frac{\rho}{m}$
\iftechreport
(derived from Corollary~\ref{cl:1.0-lower-bound}),
\else
,
\fi
which is much less than the $R_{\fmit{min}}=\theta\frac{\rho}{m+1}$
for EARDet and $R_{\fmit{min}}=\theta\frac{\rho}{m}$ for FM and AMF-FM. This is especially true
when the memory is extremely limited (i.e. $n_{\gamma} \gg m$), where
$n_{\gamma}$ is the maximum number of legitimate flows at the
threshold rate $\gamma$, and $m$ is the number of counters
for each individual detector (each \EFD in \twinEFD has $m/2$ counters).

Figures~\ref{fig:detectable-rate-1} and~\ref{fig:detectable-rate-2}
compare the $R_{\fmit{min}}$ amongst 
these three detectors given two link
capacities: 1) $\rho=10^5\gamma$ (i.e., $n_{\gamma} = 10^5$), and 2)
$\rho=10^7\gamma$ (i.e., $n_{\gamma} = 10^7$). The results suggest
that \twinEFD and \ETE have a much lower $R_{\fmit{min}}$ 
than that of other detectors
when memory is limited, and the $R_{\fmit{min}}$ is insensitive to memory size
because \EFD can add levels to overcome memory shortage.

For bursty flows, \name's  $R_{\fmit{min}}$ 
is competitive to AMF-FM, due to EARDet.

\fi

\section{Evaluation}
\label{sec:eval}

We experimentally evaluate \ETE, \EFD, EARDet, and AMF-FM
with respect to worst-case damage~\cite[Sec.~5.1]{full-paper}. 
We consider various
large-flow patterns and memory limits and assume background traffic
that is challenging for \ETE and \EFD. The experiment results confirm
that \ETE outperforms other schemes, especially when memory is
extremely limited.

\subsection{Experiment Settings}\label{eval:exp-settings}

\paragraph{Link settings.}
Since the required memory space of a large-flow detector is sublinear to
link capacity, we set the link capacity to $\rho = 1$Gbps, which is high
enough to incorporate the realistic background traffic dataset while
ensuring the simulation can finish in reasonable time. 
We choose a very low threshold rate $\gamma = 12.5$ KB/s, 
so that the number of full-use legitimate flows 
$n_{\gamma} = \rho/\gamma$ is $10000$, 
ensuring that the link is as challenging as a backbone link 
(as analyzed in 
\iftechreport
Section~\ref{ssec:efd-complexity}
\else
our technical report~\cite{full-paper}
\fi
).
The flow specification is set to 
$\func{TH}{t} = \gamma t + \beta$, where $\beta$ is set to 3028 bytes
(which is as small as two maximum-sized packets,
making bursty flows easier to catch).

The results on this 1Gbps link allow us to extrapolate 
detector performance to 
high-capacity core routers, e.g.,
in a 100Gbps link with $\gamma = 1.25$ MB/s.
Because \name's performance 
with a given number of counters
is mainly related to 
the ratio between link capacity and threshold rate
$n_{\gamma}$ (as discussed in
\iftechreport
Section~\ref{ssec:efd-detection-prob}),
\else
our technical report~\cite{full-paper}),
\fi
\name's worst-case performance will scale linearly in link capacity
when the number of counters 
and the ratio between link capacity and threshold rate is held constant.
AMF-FM, on the other hand, 
performs worse as the number of flows increases
\iftechreport
(according to Appendix~\ref{ssec:appendix-mf-analysis}
and~\ref{ssec:appendix-fm-analysis}).
\else
(according to our technical report~\cite{full-paper}).
\fi
Thus, with increasing link capacity,
AMF-FM may face an increased number of actual flows,
resulting in worse performance.
In other words, AMF-FM's worst-case damage
may be superlinear in link capacity.
As a result, if \name outperforms AMF-FM in small links,
\name will outperform AMF-FM by at least as large a ratio in larger links.

\paragraph{Background traffic.}
We consider the worst background traffic for \EFD and \name:
\iftechreport
we determine the worst-case traffic according to Theorem~\ref{th:efd-worst-case}.
\else 
we determine the worst-case traffic in our technical report~\cite[Thm.~1]{full-paper}.
\fi
Aside from attack traffic, the rest of the link capacity is completely filled with full-use legitimate flows running
at the threshold rate $\gamma = 12.5$ KB/s.
The total number of attack flows and full-use legitimate flows
is $n_{\gamma} = 10000$. Once a flow has been blacklisted
by the large-flow detectors, we fill the idle bandwidth
with a new full-use legitimate flow,
to keep the link always running
with the worst-case background traffic.

\paragraph{Attack traffic.}
We evaluate each detector against 
large flows with various average rates
$R_{\fmit{atk}}$ and duty cycle $\theta$. Their bursty period is set to be $T_b
= 0.967$s. To evaluate \EFD and \ETE against their worst-case bursty flows
($\theta T_b < 2T_c$), large flows are allotted a relatively small bursty period
$T_b = 4T_{\ell} = 0.967$s, where $T_{\ell} = \beta / \gamma = 0.242$s
 is the period of each detection level in the single \EFD.  
In \ETE, \EFD{}$^{(1)}$ uses the same detection level period
$T_{\ell}^{(1)} = T_{\ell} = 0.242$s as well.  Since  \EFD usually has $d \ge 3$ levels and $T_c \ge 3T_{\ell}$, 
it is easy for attack flows to meet $\theta T_b < 2T_c$.

In each experiment, we have $10$ artificial large flows whose
rates are in the range of $12.5$ KB/s to $12.5$ MB/s
(namely, $1$ to $1000$ times that of threshold rate $\gamma$).
The fewer large flows in the link, the longer delay
required for \EFD and \name to catch large flows;
however, the easier it is for AMF-FM to detect large flows, 
because there are fewer FPs from AMF and more frequent flow eviction in FM.
Thus, we use $10$ attack flows to challenge \name
and the results are generalizable.

\paragraph{Detector settings}
We evaluate detectors with different numbers of counters ($20 \le m \le 400$)
to understand their performance under different memory limits.
Although a few thousands of counters are available in a typical CPU, 
not all  can be used by one detector scheme.
CLEF works reasonably well with such a small number of counters 
and can perform better when more counters are available.

\begin{itemize}
\item \textbf{\emph{EARDet}}. We set the low-bandwidth threshold
to be the flow specification $\gamma t + \beta$, 
and compute the corresponding high-rate threshold, $\gamma_h = \frac{\rho}{m+1}$, for $m$
 counters as in~\cite{eardet}.

\item \textbf{\emph{\EFD}}. A \EFD has $d$ levels and $m$ counters.
We set the period of a detection level as
$T_{\ell} = \beta/\gamma = 0.242$ seconds\footnote{
If $T_{\ell} \ll \beta/\gamma$, it is hard for a large flow to reach
the burst threshold $\beta$ in such a short time;
if $T_{\ell} \gg \beta/\gamma$, the detection delay is too long,
resulting in excessive damage.}.
$d = \lfloor 1.2 \times \log_m(n) \rfloor + 1$ to have fewer flows than
the counters at the bottom level. 
The counter threshold of the bottom level is 
$\mit{TH}_{\fmit{\EFD}} = \gamma T_{\ell} + \beta$ = $2\beta$ = $6056$ Bytes.

\item \textbf{\emph{\ETE}}. 
We allocate $m/2$ counters to EARDet, and $m/4$ counters to each \EFD.
\EFD{}$^{(1)}$ and EARDet are configured like 
the single \EFD and the single EARDet above.
For the \EFD{}$^{(2)}$, we properly set its detection level period 
$T_{\ell}^{(2)}$ to guarantee detection of 
 most of bursty flows with low damage.
The details of the single \EFD and \ETE are
\iftechreport
in Table~\ref{table:efd-setting} (Appendix~\ref{ssec:appendix-table-figure}).
\else
in our technical report~\cite{full-paper}.
\fi

\item \textbf{\emph{AMF-FM}}.
We allocate half of the $m$ counters to AMF and the rest to FM.
AMF has four stages (a typical setting in~\cite{Estan2003}),
each of which contains $m/8$ counters.
All $m$ counters are leaky buckets with a drain rate of 
$\gamma$ and a bucket size $\beta$.
\end{itemize}

\iftechreport 
\begin{figure*}[!th]
\centering
\subfigure[\small{Flat large flows, $\theta = 1.0$}] {
  \label{fig:damage-flat}
  \includegraphics[width=1.25\textwidth, trim= 200 10 100 10]{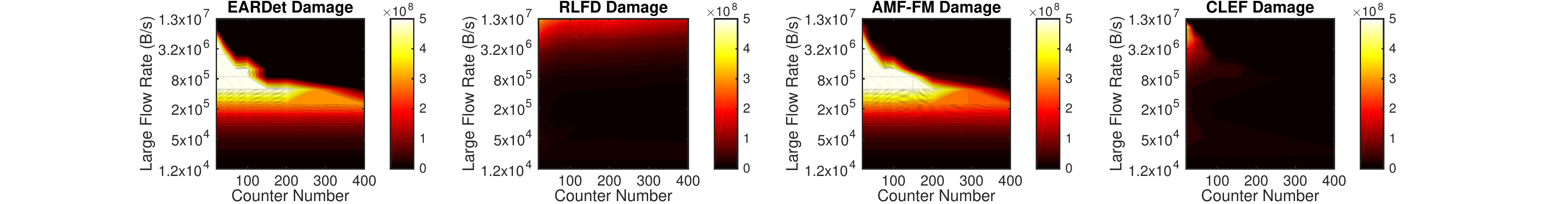}}
\subfigure[Bursty large flows, $\theta = 0.50$] {
  \label{fig:damage-burst-4}
  \includegraphics[width=1.25\textwidth, trim= 200 10 100 10]{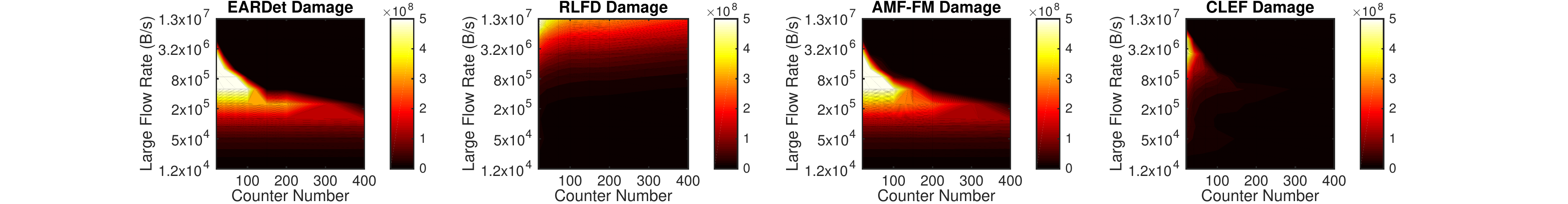}}
\subfigure[\small{Bursty large flows, $\theta = 0.25$}] {
  \label{fig:damage-burst-2}
  \includegraphics[width=1.25\textwidth, trim= 200 10 100 10]{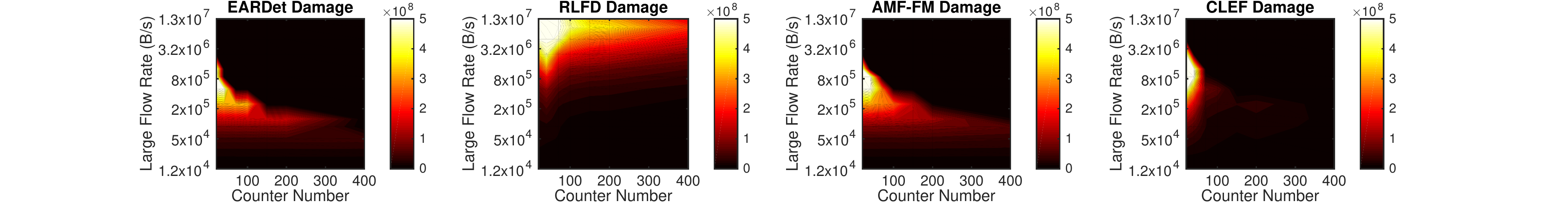}}
\subfigure[Bursty large flow, $\theta = 0.10$] {
  \label{fig:damage-burst-1}
  \includegraphics[width=1.25\textwidth, trim= 200 10 100 10]{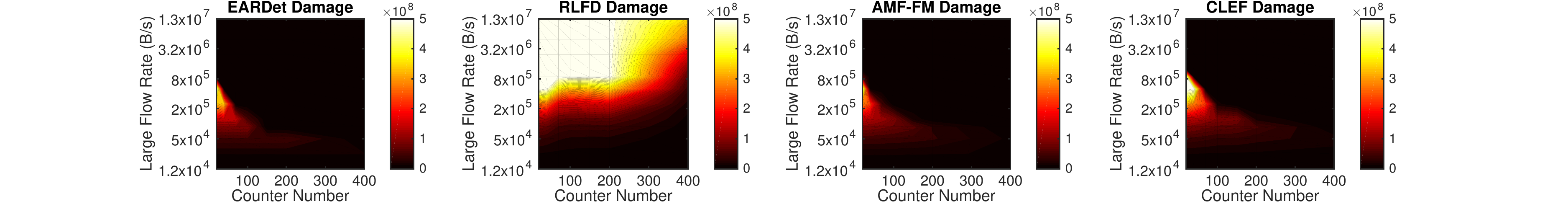}}
\subfigure[\small{Bursty large flows, $\theta = 0.02$}] {
  \label{fig:damage-burst-5}
  \includegraphics[width=1.25\textwidth, trim= 200 10 100 10]{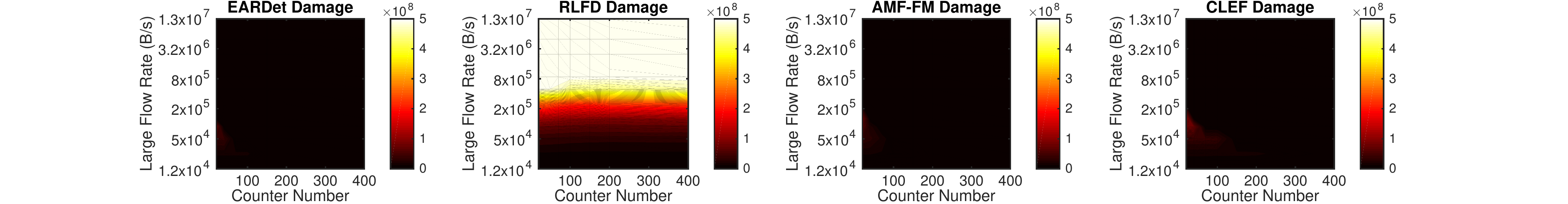}}
\vspace*{-4mm}
\caption{Damage (in Bytes) caused by $200$-second large flows 
at different average flow rate $R_{\fmit{atk}}$ (in Byte/s) and 
duty cycle $\theta$ under detection of different schemes 
with different number of counters $m$.
The larger the dark area, the lower the damage guaranteed by a scheme.
Areas with white color are damage equals or exceeds $5\times10^8$.
\ETE outperforms other schemes in detecting flat flows, and has
competitive performance to AMF-FM and EARDet over bursty flows.}
\end{figure*}

\else 

\vspace*{-6mm}
\begin{figure*}[!th]
\centering
\subfigure{
  \includegraphics[width=0.88\textwidth, trim= 90 0 540 0, clip]{fig/full-link-exp/damage-flat.eps}}
\subfigure{
  \includegraphics[width=0.90\textwidth, trim= 530 0 90 0, clip]{fig/full-link-exp/damage-flat.eps}}
\vspace*{-3mm}
\caption{Damage (in Bytes) caused by $200$-second large flows 
at different average flow rate $R_{\fmit{atk}}$ (in Byte/s) and 
duty cycle $\theta = 1.0$ (flat large flows) under detection of different schemes 
with different number of counters $m$.
The larger the dark area, the lower the damage guaranteed by a scheme.
Areas with white color are damage equals or exceeds $5\times10^8$.
Figures of bursty flows are shown in our technical report~\cite{full-paper}.
\ETE outperforms other schemes in detecting flat flows, and has
competitive performance to AMF-FM and EARDet over bursty flows.}
\label{fig:damage-flat}
\end{figure*}

\fi

\begin{figure*}[!th]
\centering
\subfigure[\small{Flat, $\theta = 1.0$}] {
  \label{fig:com-damage-flat}
  \includegraphics[width=0.3\textwidth, trim= 0 5 0 15]{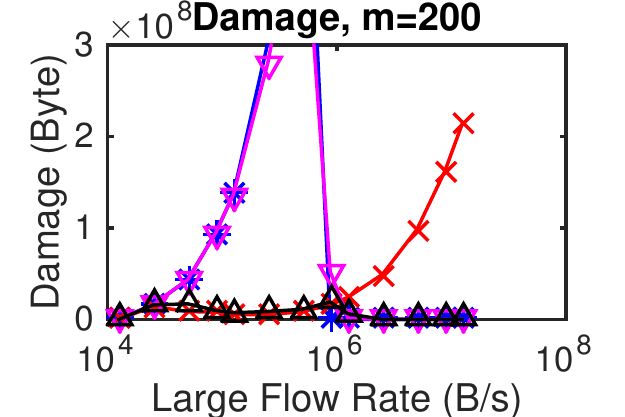}}
\subfigure[\small{Bursty, $\theta = 0.50$}] {
  \label{fig:com-damage-burst-4}
  \includegraphics[width=0.3\textwidth, trim= 0 5 0 15]{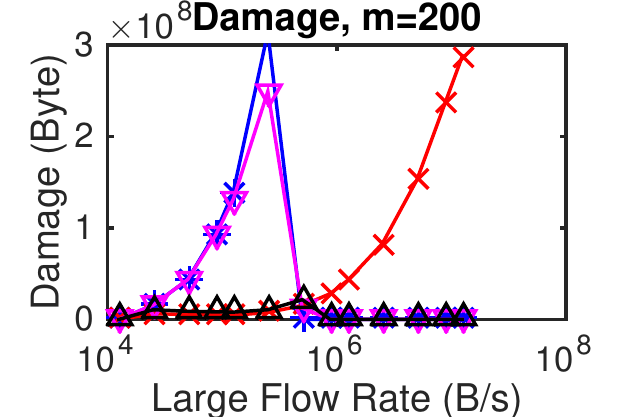}}
\subfigure[\small{Bursty, $\theta = 0.25$}] {
  \label{fig:com-damage-burst-2}
  \includegraphics[width=0.3\textwidth, trim= 0 5 0 15]{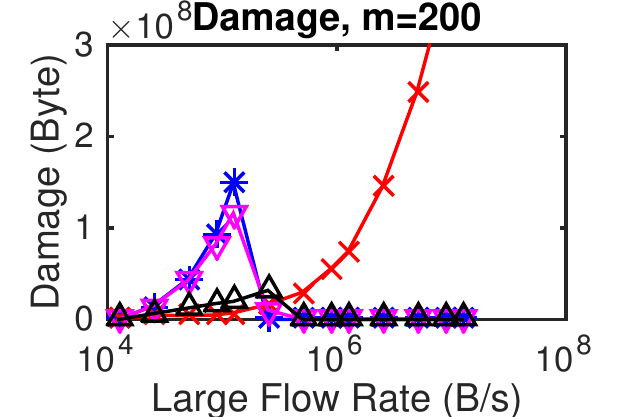}}
\subfigure[\small{Bursty, $\theta = 0.10$}] {
  \label{fig:com-damage-burst-1}
  \includegraphics[width=0.3\textwidth, trim= 0 5 0 15]{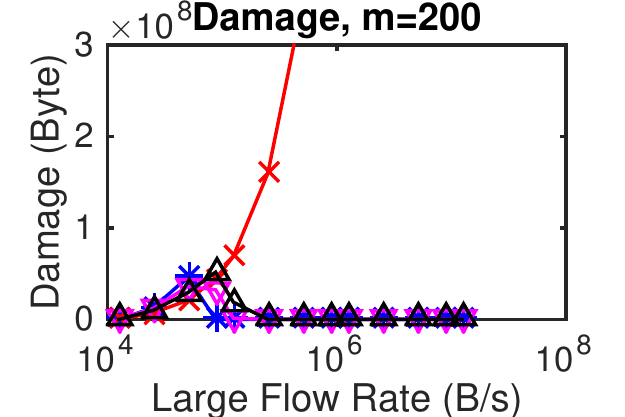}}
\subfigure[\small{Bursty, $\theta = 0.02$}] {
  \label{fig:com-damage-burst-5}
  \includegraphics[width=0.3\textwidth, trim= 0 5 0 15]{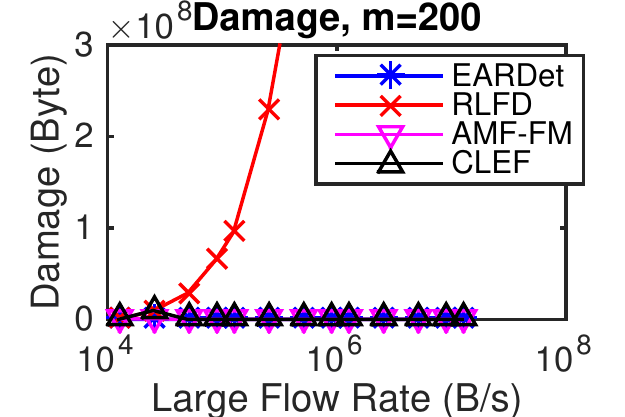}}
\vspace*{-4mm}
\caption{Damage (in Bytes) caused by $200$-second large flows at 
different average rate $R_{\fmit{atk}}$ (in Byte/s) 
and duty cycle $\theta$.
Each detection scheme uses $200$ counters in total.
The clear comparison among schemes suggests \ETE outperforms 
others with low damage against various large flows.}
\end{figure*}

\begin{figure*}[!th]
\centering
\subfigure[\small{Flat, $\theta = 1.0$}] {
  \label{fig:com-fn-flat}
  \includegraphics[width=0.3\textwidth, trim= 0 5 0 15]{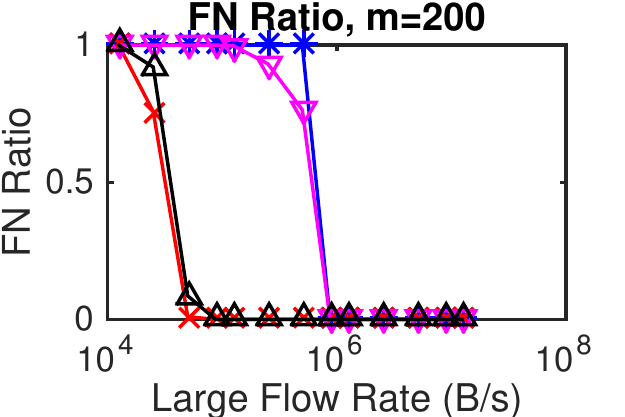}}
\subfigure[\small{Bursty, $\theta = 0.50$}] {
  \label{fig:com-fn-burst-4}
  \includegraphics[width=0.3\textwidth, trim= 0 5 0 15]{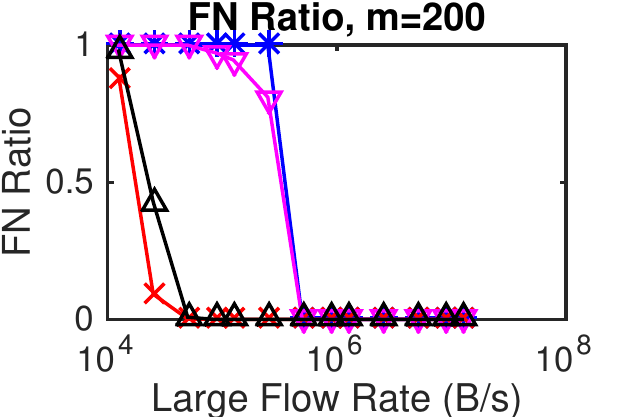}}
\subfigure[\small{Bursty, $\theta = 0.25$}] {
  \label{fig:com-fn-burst-2}
  \includegraphics[width=0.3\textwidth, trim= 0 5 0 15]{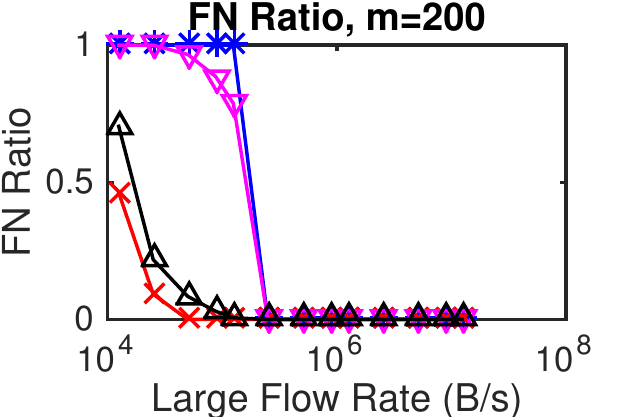}}
\subfigure[\small{Bursty, $\theta = 0.10$}] {
  \label{fig:com-fn-burst-1}
  \includegraphics[width=0.3\textwidth, trim= 0 5 0 15]{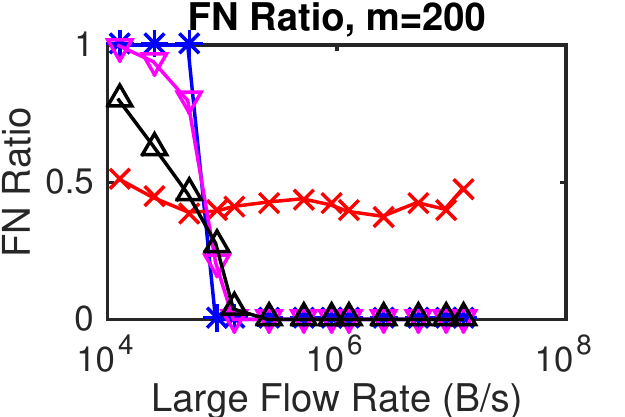}}
\subfigure[\small{Bursty, $\theta = 0.02$}] {
  \label{fig:com-fn-burst-5}
  \includegraphics[width=0.3\textwidth, trim= 0 5 0 15]{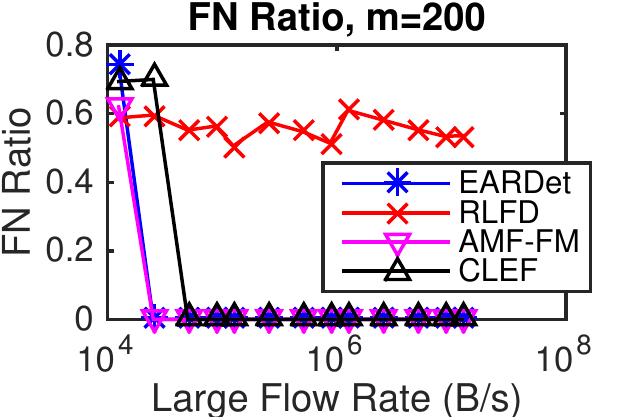}}
\vspace*{-4mm}
\caption{FN ratio in a $200$-second detection for large flows
at different average rate $R_{\fmit{atk}}$ (in Byte/s) 
and duty cycle $\theta$.
Each detection scheme uses $200$ counters in total.
\ETE is able to detect (FN$<1.0$) low-rate flows undetectable (FN$=1.0$) 
by AMF-FM or EARDet.}
\end{figure*}

\subsection{Experiment Results}
For each experiment setting (i.e., attack flow configurations
and detector settings), we did $50$ 
repeated runs and present the averaged results.

\iftechreport
Figure~\ref{fig:damage-flat} to~\ref{fig:damage-burst-5} demonstrate
\else
Figure~\ref{fig:damage-flat} demonstrates
\fi
the damage caused by large flows at different average rates, duty cycles,
and number of detector counters during $200$-second experiments; 
the lighter the color, the higher the damage.
The damage $\ge 5\times10^8$~Byte is represented by the color white.
Figures~\ref{fig:com-damage-flat} to~\ref{fig:com-damage-burst-5} compare
 damage in cases of different detectors with $200$ counters.
Figures~\ref{fig:com-fn-flat} to~\ref{fig:com-fn-burst-5} show 
the percentage of FNs produced 
by each detection scheme with $200$ counters within $200$ seconds.
We cannot run infinitely-long experiments to show the $+\infty$ damage
produced by detectors like EARDet and AMF-FM over low-rate flows,
so we use the FN ratio to suggest it here. An FN of $1.0$ means that the detector
fails to identify  any large flow in $200$ seconds and is likely 
to miss large flows in the future. Thus, an infinite damage is assigned.
On the contrary, if a detector has FN rate $< 1.0$, it is able to detect 
remaining large flows at some point in the future.

\paragraph{\name ensures low damage against flat flows.}
Figures \ref{fig:damage-flat}, \ref{fig:com-damage-flat},
and~\ref{fig:com-fn-flat} 
\iftechreport
support our theoretical analysis (in Section~\ref{sec:analysis}) 
\else
show
\fi
that \EFD and \ETE
work effectively at detecting low-rate flat large flows and guaranteeing low
damage. On the contrary, such flows cause much higher damage against EARDet and
AMF-FM.  The nearly-black figure (in Figure~\ref{fig:damage-flat}) for \ETE
shows that \ETE is effective for both high-rate and low-rate flat flows with different memory limits.  Figure~\ref{fig:com-damage-flat} shows a clear damage
comparison among detector schemes. 
\ETE, EARDet, and AMF-FM all limit the damage
to nearly zero for high-rate flat flows. 
However, the damage limited by \ETE is
much lower than that limited by AMF-FM and EARDet for the low-rate flat flows.
EARDet and AMF-FM results show a sharp top boundary
that reflects the damage dropping to zero
at the guaranteed-detection rates.

The damage limited by an individual \EFD is proportional to the
large-flow rate when the flow rate is high.
Figure~\ref{fig:com-fn-flat} suggests that AMF-FM and EARDet are 
unable to catch most low-rate flat flows ($R_{\fmit{atk}} < 10^6$ Byte/sec),
which explains the high damage by low-rate 
flat flows against these two schemes.
\iftechreport
This supports our theoretical analysis of AMF-FM and EARDet 
in Table~\ref{table:th-compare}: the infinite damage by low-rate flows
against AMF-FM and EARDet.
\fi

\paragraph{\name ensures low damage against various bursty flows.}
Figures~\ref{fig:com-damage-burst-4} to~\ref{fig:com-damage-burst-5}
demonstrate the damage caused by bursty flows with different duty cycle 
$\theta$. The smaller the $\theta$ is, 
the burstier the flow. As the large flows become burstier,
the EARDet and AMF-FM schemes improve 
at detecting flows whose average rate is low. 
Because the rate at the burst is $R_{\fmit{atk}}/\theta$, which increases as $\theta$ decreases, 
thus EARDet and AMF-FM are able to detect these flows 
even though their average rates are low.
For a single \EFD, the burstier the flows are, the harder it becomes to detect
the large flows and limit the damage. As we discussed in 
Section~\ref{ssec:eardet-efd-hybrid}, when the burst duration
$\theta T_b$ of flows is smaller 
than the \EFD detection cycle $T_c$, a single \EFD
has nearly zero probability of detecting such attack flows. Thus, we
need \twinEFD in \ETE to detect bursty flows missed by EARDet
in \ETE, so that  \ETE's damage is still low as the figures show.
When the flow is very bursty (e.g., $\theta \le 0.1$),
the damage limitation of the \ETE scheme is dominated by EARDet.

Figures~\ref{fig:com-damage-burst-4} to~\ref{fig:com-damage-burst-5}
present a clear comparison among different schemes against bursty flows.
The damage limited by \ETE is lower than that limited by AMF-FM and
 EARDet, when $\theta$ is not too small (e.g., $\theta \ge 0.25$). 
Even though  AMF-FM and EARDet have lower damage for very bursty flows
(e.g., $\theta \le 0.1$) than the damage limited by \ETE,
the results are close 
because  \ETE is assisted by an EARDet with $m/2$ counters.
Thus, \ETE guarantees a low damage limit for a wider range
of large flows than the other schemes.

\paragraph{\name outperforms others in terms of FN and FP.}
To make our comparison more convincing,
we examine schemes with classic metrics: FN and FP.
Since we know all four schemes have no FP, we simply check the FN ratios
in Figures~\ref{fig:com-fn-flat} to~\ref{fig:com-fn-burst-5}.
Generally, \name has a lower FN ratio than AMF-FM and EARDet do.
\name can detect large flows at a much lower rate with zero FN ratio,
and is competitive to AMF-FM and EARDet against very bursty flows
(e.g., Figures~\ref{fig:com-fn-burst-4} and~\ref{fig:com-fn-burst-5}).

\paragraph{\name is memory-efficient.}
Figure~\ref{fig:damage-flat} shows
that the damage limited by \EFD is relatively insensitive
to the number of counters.
This suggests that \EFD can work with limited memory and is scalable
to larger links without requiring a large amount of high-speed memory. 
This can be explained by \EFD's recursive subdivision, 
by which we simply add one or more levels when the memory limit is low.
Thus, we choose \EFD to complement EARDet in \ETE.

In Figure~\ref{fig:damage-flat}, \name ensures a low damage 
(shown in black) with tens of counters, 
while AMF-FM suffers from a high damage (shown in light colors),
even with $400$ counters. 
\iftechreport
This supports our theoretical results in Figures~\ref{fig:detectable-rate-1}
and~\ref{fig:detectable-rate-2}.
\fi

\vspace*{-4mm}
\begin{figure}[!th]
\centering
\subfigure[Damage] {
  \label{fig:ee-damage}
  \includegraphics[width=0.35\textwidth]{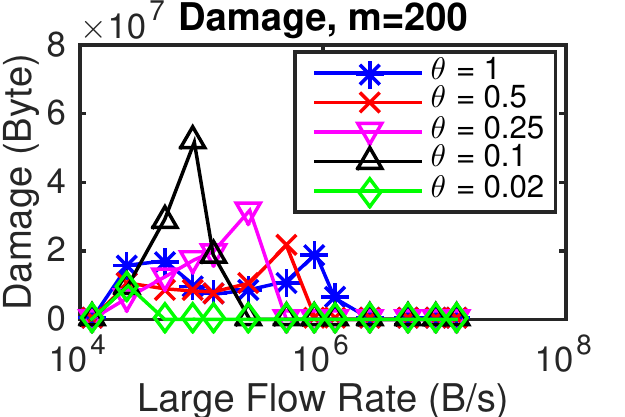}}
\subfigure[FN Ratio] {
  \label{fig:ee-fn}
  \includegraphics[width=0.35\textwidth]{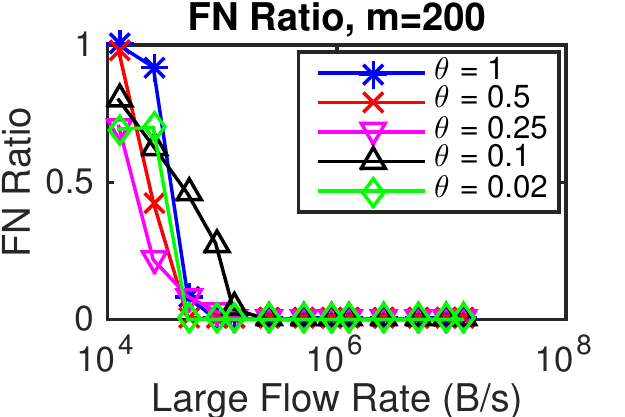}}
\vspace*{-4mm}
\caption{Damage and FN ratio for large flows at different average
rate $R_{\fmit{atk}}$ (in Byte/s) 
and duty cycle $\theta$ under detection of 
\ETE with $m=200$ counters. \ETE is insensitive to bursty flows 
across duty cycles: 1) the damages are around the same scale 
(not keep increasing as duty cycle decrease, because of EARDet), 
2) the FN ratios are stable and similar.}
\end{figure}
\vspace*{-2mm}

\paragraph{\name is effective against various types of bursty flows.}
Figures~\ref{fig:ee-damage} and~\ref{fig:ee-fn} demonstrate the changes of damage
and FN ratio versus different duty cycles $\theta$ when \ETE is used to detect
bursty flows.  In the $200$-second evaluation, as $\theta$ decreases, the
maximum damage across different average flow rates increases first by ($\theta \ge
0.1$) and then decreases by ($\theta < 0.1$). The damage increases when $\theta \ge
0.1$ because \twinEFD (in \ETE) gradually loses its capability to  detect
bursty flows. The damage therefore increases due to the increase in detection delay.

However, the maximum damage does not increase all the way as $\theta$
decreases, because when $\theta$ is getting smaller, EARDet is able to catch
bursty flows with a lower average rate. This explains the lower damage from large flows
in the $200$-second timeframe.  Figure~\ref{fig:ee-fn} shows that the FN ratio curve
changes within a small range as  $\theta$ decreases, which also indicates the
stable performance of \ETE against various bursty flows.  Moreover, the FN
ratios are all below $1.0$, which means that \ETE can eventually catch large
flows, whereas EARDet and AMF-FM cannot.

\paragraph{\name operates at high speed.}
We also evaluated the performance of a Golang-based 
implementation under real-world traffic trace 
from the CAIDA~\cite{caida} dataset.
The implementation is able to process 11.8M packets per second,
which is sufficient for a 10~Gbps Ethernet link,
which has a capacity of 14.4M packets per second.

\section{Related Work}
\label{sec:related}

The most closely related large-flow detection algorithms are described
in Section~\ref{ssec:existing-algorithms} and compared in
\iftechreport
Sections~\ref{sec:analysis} and~\ref{sec:eval}.
\else
Section~\ref{sec:eval} and further in our technical report~\cite{full-paper}.
\fi
This section discusses other related schemes.

\paragraph{Frequent-item finding.}
Algorithms that find frequent items in a stream can be applied to
large-flow detection. For example, Lossy Counting~\cite{Manku2002}
maintains a lower bound and an upper bound of each item's count. It
saves memory by periodically removing items with an upper bound below
a threshold, but loses the ability to catch items close to the
threshold. However, the theoretical memory lower bound of one-pass
exact detection is linear to the number of large flows, which is
unaffordable by in-core routers. By combining a frequent-item finding
scheme with \EFD, \name can rapidly detect high-rate large flows and
confine low-rate large flows using limited memory.

\paragraph{Collision-rich schemes.}
To reduce memory requirement in large-flow utilization, a common
technique is hashing flows into a small number of bins. However,
hash collisions may cause FPs, and FPs increase as the available
memory shrinks.
For example, both multistage filters~\cite{Estan2003, Estan2003a} and
space-code Bloom filters~\cite{kumar2006space} suffer from high FPs
when memory is limited.

\paragraph{Sampling-based schemes.}
Sampling-based schemes estimate the size of a flow based on sampled
packets. However, with extremely limited memory and thus a low
sampling rate, neither packet sampling (e.g., Sampled
Netflow~\cite{Netflow}) nor flow sampling (e.g., Sample and
Hold~\cite{Estan2003} and Sticky Sampling~\cite{Manku2002}) can
robustly identify large flows due to insufficient information. 
In contrast, \EFD in \name progressively narrows down the candidate
set of large flows, thereby effectively confining the damage caused by
large flows.

\paragraph{Top-k detection.}
Top-k heavy hitter algorithms can be used to identify flows that use
more than $1/k$ of bandwidth. Space Saving~\cite{metwally2005efficient} finds
the top-k frequent items by evicting the item with the lowest counter
value. HashPipe~\cite{sivaraman2017heavy} improves upon Space Saving
so that it can be practically implemented on switching hardware.
However, HashPipe still requires keeping 80KB to detect large flows
that use more than 0.3\% of link capacity, whereas \name can enforce flow
specifications as low as $10^{-6}$ of the link capacity using only
10KB of memory. Tong~\etal~\cite{tong2016high} propose an efficient
heavy hitter detector implemented on FPGA but the enforceable flow
specifications are several orders looser than \name. Moreover,
misbehaving flows close to the flow specification can easily bypass
such heavy hitter detectors. The FPs caused by heavy
hitters prevent network operators from applying strong punishment to
the detected flows.

Chen~\etal~\cite{chen2016counter} and Xiao~\etal~\cite{xiao2015hyper}
propose memory-efficient algorithms for estimating per-flow
cardinality (e.g., the number of packets). These algorithms, however,
cannot guarantee large-flow detection in adversarial environments due
to under- or over-estimation of the flow size.

Liu~\etal~\cite{Liu2016} 
propose a generic network monitoring framework called UniMon that allows 
extraction of various flow statistics.
It creates flow statistics for all flows,
but has high FP and FN when used to detect large flows.

\section{Conclusion}
\label{sec:conclusion}

In this paper we propose new efficient large-flow detection algorithms.
First, we develop a randomized Recursive Large-Flow Detection
(\EFD) scheme, which uses very little memory
yet provides eventual detection of persistently large flows.
Second, we develop \name, which
scales to Internet core routers and is resilient against worst-case
traffic. None of the prior approaches can achieve the same level of
resilience with the same memory limitations.
To compare attack resilience among various detectors, we
define a damage metric that summarizes the impact of attack
traffic on legitimate traffic.
\name can confine damage even when faced with the worst-case background
traffic because it combines a deterministic EARDet for the rapid detection of
very large flows and two \EFD{}s to detect near-threshold large flows.
We 
\iftechreport
proved
\else
denomstrated
\fi
that \name is able to guarantee low-damage large-flow
detection against various attack flows with limited memory,
outperforming other schemes even with \name's worst-case
background traffic. Further experimental evaluation
confirms the findings of our
theoretical analysis and shows that \name has the lowest worst-case
damage among all detectors and consistently low damage over a wide
range of attack flows.

\section{Acknowledgments}

We thank Pratyaksh Sharma and Prateesh Goyal for early work on this
project as part of their summer internship at ETH in Summer 2015.
We also thank the anonymous reviewers, whose feedback
helped to improve the paper.

The research leading to these results has received funding from the
European Research Council under the European Union's Seventh Framework
Programme (FP7/2007-2013), ERC grant agreement 617605,
the Ministry of Science and Technology of Taiwan under grant number MOST 107-2636-E-002-005,
and the US National Science Foundation under grant numbers
CNS-1717313 and CNS-0953600. We also
gratefully acknowledge support from ETH Zurich and from the Zurich
Information Security and Privacy Center (ZISC).

\bibliographystyle{splncs04}
\bibliography{paper}

\iftechreport
    \newpage
    \appendix
\section{Additional Details For \EFD Data Structure and Optimization}
\label{ssec:appendix-optimization}

\subsection{Analysis for No-FP Guarantee}
\label{ssec:appendix-no-fp}
To guarantee no FP, we only identify large flows whose
counter has no second flow, i.e. no flow hash collision $C_{\fmit{free}}$. 
If we randomly hash flows into counters
at the bottom level $L_d$, the no-collision probability for a counter is
$Pr(C_{\fmit{free}}) = [\frac{m-1}{m}]^{n_d - 1}$,
where $n_d$ is the number of flows selected into $L_d$.
Because we want to have $d$ as small as possible, thus, we usually may choose
$d = \lceil \log_m n \rceil$, where n is the total number of flows in the link.
Thus, $n_d \le m$ on average. Thus,

\begin{small}  
\begin{equation}
Pr(C_{\fmit{free}}) = \bigg[\frac{m-1}{m}\bigg]^{m\frac{n_d - 1}{m}}
                    \approx e^{-\frac{n_d - 1}{m}}
\end{equation}  
\end{small}

\noindent
When $n_d \approx m$, the no-collision probability
$Pr(C_{\fmit{free}}) \approx \frac{1}{e} = 0.368$,
which gives a collision probability for each flow of $0.632$.

To avoid the high collision probability in the regular hash above,
we randomly pick $m$ flows (out of $n_d$ flows) instead.
Each of $m$ flows is monitored by a dedicated counter 
(which does not introduce additional FNs, because $n_d \le m$).
To efficiently implement this counter assignment,
we can use Cuckoo hashing~\cite{pagh2001cuckoo} to achieve constant 
expected flow insertion time and 
worst-case constant lookup and update time.
Cuckoo hashing resolves collisions by using
two hash functions instead of only one in regular hashing.
As in~\cite{mitzenmacher2009some}, Mitzenmacher shows that, 
with three hash functions, Cuckoo hashing
can achieve expected constant insertion and lookup time 
with load factor of $91\%$. Thus, when $m = n_d$, 
the Cuckoo hashing can achieve $Pr(C_{\fmit{free}}) \approx 0.91$, 
which is still much larger than $Pr(C_{\fmit{free}}) \approx 0.368$
in the regular hashing. As $n_d$ is usually less than $m$ (because
we set $d$ to be the ceiling of $\log_m n$), 
it is reasonable to treat the $Pr(C_{\fmit{free}}) \approx 1$ 
in our later analysis. Cuckoo hashing requires to store both the
key ($48$ bits for IPv4, $144$ bits for IPv6) and value ($32$ bits) 
of an entry, thus, for each counter, 
we need space for the flow ID and the counter value.

\subsection{Shrinking Counter Entry Size}
\label{ssec:appendix-shrink-counter-size}
As we discussed, the number of flows hashed into the bottom level
is much less than $m$ (e.g. at most $2^{10}$).
a key space of $96$ bits ($288$ for IPv6) is too large for less than $2^{10}$  
keys. We can hash the flow IDs into a smaller key space, e.g. $48$ bits to 
save memory size. For each flow,
although hash collision could happen and may result in FP in
the detection in the bottom level, the probability is less than 
$1 - [\frac{2^{48}-1}{2^{48}}]^{2^{10}-1} \approx 2^{-38}$
which is very small. For systems can tolerate 
such extremely low FP probability, we recommend it to do so.

\section{Additional Analysis}
\subsection{Flow Memory Analysis}\label{ssec:appendix-fm-analysis}
We analyze the Flow Memory (FM) with random flow eviction mechanism,
which is applied with multistage filters in~\cite{Estan2003}.
For each incoming packet whose flow is not tracked, such FM randomly picks a
flow from the tracked flows and the new flow to evict.
Thus, for each packet of the flow not tracked, the existing tracked
flow has a probability $P_{e} = \frac{1}{m+1}$ to be evicted,
where $m$ is the number of counters in the FM.

\begin{theorem}\label{th:fm-least-rate}
In a link with total traffic rate of $R$ ($\le \rho$), 
the packet size of $S_{\fmit{pkt}}$,
and the large-flow threshold $\func{TH}{t} = \gamma t + \beta$,
a Flow Memory with $m$ counters is able to detect large flows at rate
around or higher than $\frac{\beta}{S_{\fmit{pkt}}}\frac{R}{m}$
with high probability. 
\end{theorem}

\paragraph{Proof sketch:}
We assume number of packets arriving at the FM
per second is at the packet rate of $R_{\fmit{pkt}}$,
thus the time gap between two incoming packets is 
$T_{\fmit{pkt}} = \frac{1}{R_{\fmit{pkt}}} = \frac{S_{\fmit{pkt}}}{R}$.
For a newly tracked flow $f$
at time stamp $0$, the $k$th eviction happens at $k\cdot T_{\fmit{pkt}}$,
and $P_{e} = \frac{1}{m+1}$ is the probability
that flow $f$ is evicted at the $k$th eviction.
Evictions are not triggered by packets of flows  
being tracked, however the number of flows untracked is far larger than 
the number of flows being tracked, thus we can approximate treat
the time gap between evictions as $T_{\fmit{pkt}}$.
Thus, the expected time length for the flow $f$ to be tracked is 

\begin{small}
\begin{equation}\label{eq:fm-track-time}
\begin{aligned}
    &E(T_{\fmit{track}})\\
    &=\sum_{k=1}^{+\infty} P_e (1 - P_e)^{k-1} kT_{\fmit{pkt}}\\
    &=\lim_{k\to+\infty}(1-P_e) \bigg( \frac{1 - (1-P_e)^k}{P_e} - 
        (k+1)(1 - P_e)^k\bigg) T_{\fmit{pkt}}\\
    &=\frac{1-P_e}{P_e} T_{\fmit{pkt}} = m\cdot T_{\fmit{pkt}}
\end{aligned}
\end{equation}
\end{small}

As the FM uses leaky bucket counters to enforce the large-flow threshold
$\frac{TH}{t}=\gamma t + \beta$ 
(defined in Section~\ref{ssec:large-flow-detection}), 
the counter threshold is the burst threshold $\beta$.
Thus, to detect a large flow at traffic rate of $R_{atk}$,
the FM requires the large flow being tracked at least for a time of
$\beta / R_{atk}$, otherwise the counter value cannot reach the threshold. 
Therefore,

\begin{small}
\begin{equation}\label{eq:fm-least-rate}
\begin{aligned}
    R_{atk} &> \frac{\beta}{E(T_{\fmit{track}})}
            =\frac{\beta}{m\cdot T_{\fmit{pkt}}}
            = \frac{\beta}{S_{\fmit{pkt}}}\frac{R}{m}
\end{aligned}
\end{equation}
\end{small}
\noindent
Thus for the large flows at rates far smaller than the 
$\frac{\beta}{S_{\fmit{pkt}}}\frac{\rho}{m}$ are likely to be evicted before
violating the threshold $\beta$.$\blacksquare$

In the practice, the packet size is not fixed, but we treat it with fixed size
for analyzing the least $R_{atk}$ changes along with the $m$.
Because the real packet size is also limited in $1514$ Bytes,
the $\frac{\beta}{S_{\fmit{pkt}}}$ is a bounded factor.
As the $\beta$ is usually larger than the maximum packet size,
the $\frac{\beta}{S_{\fmit{pkt}}} > 1$ for sure.

We can see the scale of the large flow rate can be detected by FM
is similar to that can be detected by EARDet (i.e., $\frac{\rho}{m+1}$, 
where $\rho$ is the link capacity). They both increase as $\frac{1}{m}$
increases. In the worst case of the FM, when the traffic rate is
at link capacity ($R=\rho$), the least detectable average rates 
$R_{\fmit{atk}}$ of the FM and the EARDet are at the same scale.
One difference between them is that the EARDet can guarantee 
deterministic detection, while the Flow Memory detects flows 
probabilistically. Our simulations in Section~\ref{sec:eval} support
the analysis above.

\subsection{Multistage Filter Analysis}\label{ssec:appendix-mf-analysis}
According to the theoretical analysis in~\cite{Estan2003},
a $m$-counter multistage filter with $d$ stages each of which 
has $m/d$ counters, the probability for a flow hashed 
into a counter in each stage without
collision ($C_{\fmit{free}}$) to other flows is as follows. We let $m' = m/d$,
and assume there are $n$ flows in total, then

\begin{small}
\begin{equation}\label{eq:mf-no-collision}
\begin{aligned}
    Pr(C_{\fmit{free}}) &= 1 - (1 - (1-\frac{1}{m'})^{n-1})^d\\
        &=1 - (1 - (1-\frac{1}{m'})^{m'\frac{n-1}{m'}})^d\\
        &\approx 1 - (1 - e^{-\frac{n-1}{m'}})^d\\
        &\rightarrow 0 \mit{, when $n \rightarrow +\infty$}
\end{aligned}
\end{equation}
\end{small}

\noindent
where we assume the $m' \gg 1$ and $n/m' \gg 1$.
The assumptions are reasonable: 1) the number of counters $m$ is usually
around hundreds, and the $d$ is typically chosen as $4$ in~\cite{Estan2003},
therefore $m' \gg 1$;
2) we aim to use very limited counters to detect large flows from a large
number of legitimate flows, thus $n/m' \gg 1$.

In the case that every legitimate flow is higher than
the half of the threshold rate $\gamma/2$, the false positive rate is 
almost $100\%$, because the $Pr(C_{\fmit{free}})$ is close to $100\%$.
Any collision in a counter results in that the counter value violates 
the counter threshold and thus a falsely positive on legitimate flows.

\subsection{\EFD Worst-case Background Traffic}
\label{ssec:appendix-worst-efd}

\paragraph{General case: weighted balls-into-bins problem.}
In the well-known balls-into-bins problem, 
we have $m$ bins and $n$ balls. 
For each ball, we randomly throw it into one of $m$ bins.

We treat the flows in the network as the balls, 
and the counter array as the bins.
Hashing flows into counters is just like randomly throwing balls into bins,
where each flow is a weighted ball with weight of its traffic volume sent
during a period $T_{\ell}$ of each level $L_k$ ($1 \le k \le d$).

\paragraph{Worst case: single-weight balls-into-bins problem}
We assume the rate threshold $\gamma$ of our flow specification, 
$\func{TH}{t} = \gamma t + \beta$, is $\gamma = \frac{\rho}{N}$,
where the $\rho$ is the outbound link capacity.
In the general case, the legitimate flows are at average rates 
less than or equal to the threshold rate $\gamma$, however we show 
that the worst case background traffic for \EFD to detecting a large flow is 
that all legitimate flows are sending
traffic at the rate of the threshold rate $\gamma$ 
(Theorem~\ref{th:efd-worst-case}).
As the inbound link capacity can be larger than the outbound one, 
there still could be attack flows in this case.
We prove the Theorem~\ref{th:efd-worst-case} 
by the Theorem~\ref{th:weighted-ball-in-bin} 
from Berenbrink~\etal~\cite{berenbrink2008weighted}
which is for weighted balls-into-bins games.

\begin{theorem}\label{th:weighted-ball-in-bin}
\textbf{Berenbrink~\etal's Theorem 3.1} 
For two weighted balls-into-bins games $B(w,n,m)$ and $B'(w',n,m)$ of $n$ balls
and $m$ bins, the vectors $w=(w_1, ..., w_n)$ and $w'=(w'_1, ..., w'_n)$ 
represent the weight of each ball in two $B$ and $B'$, respectively.
If $W=\sum^n_{i=1} w_i = \sum^n_{i=1} w'_i$ and 
$\sum^k_{i=1} w_i \ge \sum^k_{i=1} w'_i$ for all $1 \le k \le n$,
then $E[S_i(w)] \ge E[S_i(w')]$ for all $1 \le i \le m$, 
where the $S_i(w)$ is the total load of the $i$ highest bins, 
and the $E(S_i(w))$ is the expected $S_i(w)$ across all $m^n$ possible
balls-into-bins combinations.
\end{theorem}

\paragraph{Lemma~\ref{lm:efd-worst-case} and Proof sketch}

\begin{lemma}\label{lm:efd-worst-case}
The \EFD has the lowest probability to correctly select 
the counter of a large flow $f_{\fmit{atk}}$ to the next level, 
when the legitimate flows use up all legitimate bandwidth.
\end{lemma}

We assume $\mathcal{C}_1$ and $\mathcal{C}_2$ are two different
counter states after adding the attack traffic and 
the traffic of some legitimate flows,
and there are $V$ more volume of traffic allowed to send by 
the other legitimate flows before the total volume of legitimate flows
reaches the outbound link capacity.
Let $V_{\fmit{atk}}$ be the value of the counter assigned to $f_{\fmit{atk}}$,
and the $V_{\fmit{max}}$ be the maximum value of other counters.
In the $\mathcal{C}_1$, we let $V_{\fmit{atk}} > V_{\fmit{max}} + V$;
in the $\mathcal{C}_2$, we let $V_{\fmit{atk}} \le V_{\fmit{max}} + V$.
Hence, $\mathcal{C}_1$ and $\mathcal{C}_2$ cover all possible counter states.
As there are still up to $V$ volume of legitimate flows
can be added into counters.
We use $V'_{\fmit{atk}}$ and $V'_{\fmit{max}}$ to represent the final value of
$V_{\fmit{atk}}$ and $V_{\fmit{max}}$.
Thus, the probability to select the counter of $f_{\fmit{atk}}$ is 

\begin{small}
\begin{equation}\label{eq:efd-worst-case-1}
    Pr(V'_{\fmit{atk}} > V'_{\fmit{max}}) 
        = Pr(V'_{\fmit{atk}} > V'_{\fmit{max}}|\mathcal{C}_1)Pr(\mathcal{C}_1)
        + Pr(V'_{\fmit{atk}} > V'_{\fmit{max}}|\mathcal{C}_2)Pr(\mathcal{C}_2)
\end{equation}
\end{small}

\noindent
Because $V_{\fmit{atk}} > V_{\fmit{max}} + V$ in $\mathcal{C}_1$, 
and the $V'_{\fmit{max}}$ cannot exceed $V_{\fmit{max}} + V$, 
thus always $V'_{\fmit{atk}} > V'_{\fmit{max}}$. Then,

\begin{small}
\begin{equation}\label{eq:efd-worst-case-2}
    Pr(V'_{\fmit{atk}} > V'_{\fmit{max}}) = Pr(\mathcal{C}_1)
        + Pr(V'_{\fmit{atk}} > V'_{\fmit{max}}|\mathcal{C}_2)Pr(\mathcal{C}_2)
\end{equation}
\end{small}

Let $x$ be the amount of legitimate traffic
added into counters after $\mathcal{C}_2$, where $0 \le x \le V$.
If the $x = V$, then there is a chance to have all 
$V$ added on the $V_{\fmit{max}}$, 
and thus $V'_{\fmit{max}} = V_{\fmit{max}} + V = V_{\fmit{atk}} = V'_{\fmit{atk}}$,
so that $Pr(V'_{\fmit{atk}} > V'_{\fmit{max}}|\mathcal{C}_2)$ is lower
than that when $x < V$.
Therefore, $Pr(V'_{\fmit{atk}} > V'_{\fmit{max}})$ is lower
in the case that legitimate flows fully use the link capacity than other cases.
Thus, the Lemma~\ref{lm:efd-worst-case} is proved.
$\blacksquare$

\paragraph{Proof sketch of Theorem~\ref{th:efd-worst-case}}
We first just consider the legitimate traffic but not the attack flow.
As Lemma~\ref{lm:efd-worst-case} illustrated, 
the more traffic sent from legitimate flows,
the harder for \EFD to correctly select the counter with 
the attack flow $f_{\fmit{atk}}$, 
thus to have the worst \EFD detection probability, 
legitimate flow should use all outbound link capacity,
and it requires the flow number $n \ge \rho/\gamma$.

Given $n \ge \rho/\gamma$ and $m$, we first construct a
legitimate flow configuration $B(w, n, m)$, $w_i=\gamma$ 
for $1 \le i \le \rho/\gamma$ and $w_i = 0$ for $i > \rho/\gamma$
which is the worst-case legitimate configuration we want to prove,
because there are actually only the first $\rho/\gamma$ flows 
with non-zero rate.
For any legitimate flow configuration $B'(w',n,m)$
with constraint of $\sum^k_{i=1} w'_i = \rho$.
It is easy to find $\sum^k_{i=1} w_i \ge \sum^k_{i=1} w'_i$.

Thus, according to Theorem~\ref{th:weighted-ball-in-bin}
~\cite{berenbrink2008weighted}
the $E[S_i(w)] \ge E[S_i(w')]$ for all $1 \le i \le m$,
where $E[S_i(w)]$ is the expected total counter value of the $i$ highest
counters in the case of $B(w, n, m)$ and $E[S_i(w')]$ is the one in 
the case of any other legitimate flow configuration $B'(w',n,m)$.

It is not hard to find that 
the $E[S_i(w)] \ge E[S_i(w')]$ for all $1 \le i \le m$ suggests that
the variation of expected counter values across all counters of the
$B(w, n, m)$ is larger than that of the $B'(w', n, m)$.
Let $V_{\fmit{max}}$ be the maximum counter value, 
and $V_{\fmit{atk}}$ be the value of the counter randomly assigned to 
the attack flow $f_{\fmit{atk}}$ 
($V_{\fmit{atk}}$ does not count the traffic of $f_{\fmit{atk}}$). 
The higher the variation, 
the larger the expected $V_{\fmit{max}} - V_i$,
thus the harder for \EFD to correctly select the counter of $f_{\fmit{atk}}$
for the next level.

Therefore, the $B(w, n, m)$ is the worst legitimate flow configuration 
for \EFD to detect large flows.
$\blacksquare$

\begin{figure}[!th]
\centering
\includegraphics[width=0.8\textwidth]{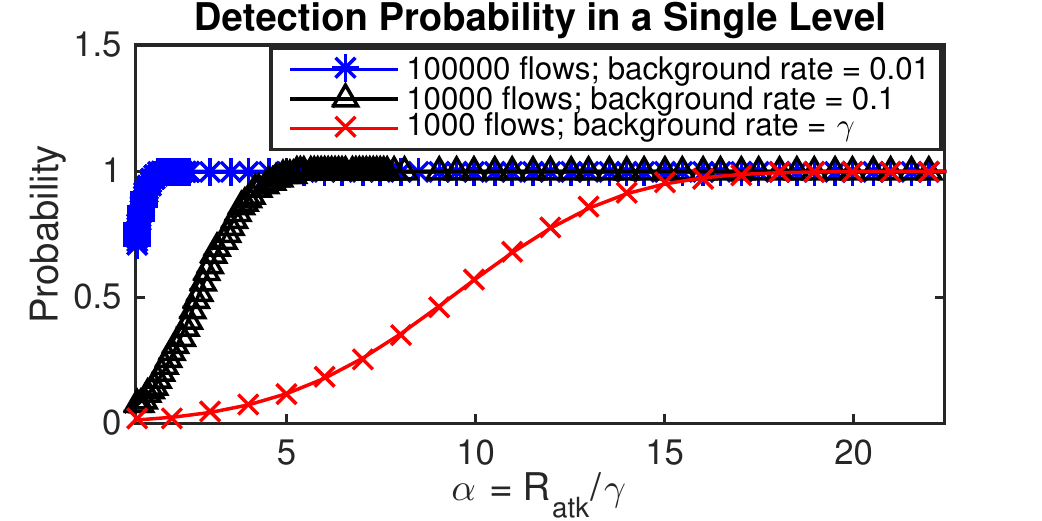}
\caption{\EFD's single-level detection probability of the 1st level
against a large flow at different rate 
$R_{\fmit{atk}} = \alpha\gamma$, when background legitimate flows
at various rates ($0.01\gamma$, $0.1\gamma$, and $\gamma$) fully use
the link capacity of $1000\gamma$. The \EFD suffers the lowest
detection probability when the legitimate flows 
are at the threshold rate $\gamma$.
}
\label{fig:worst-background}
\end{figure}

\subsection{Numeric Analysis For \EFD Detection Probability}
\label{ssec:appendix-efd-detection-prob}

\begin{figure*}[!th]
\centering
\subfigure[$n = 50$, $m = 100$] {
  \label{fig:bib:n-50-m-100}
  \includegraphics[width=0.3\textwidth]{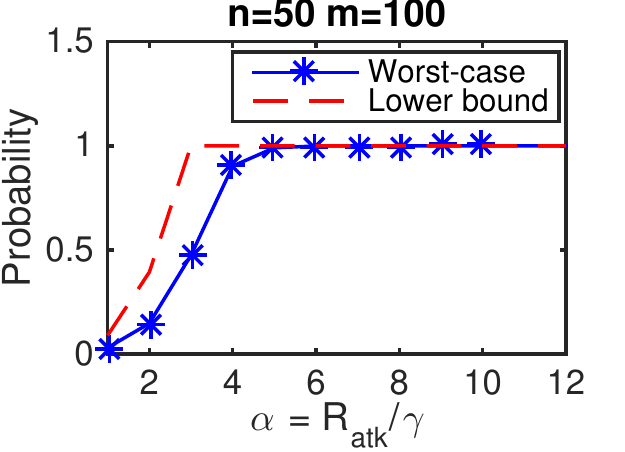}}
\subfigure[$n = 100$, $m = 100$] {
  \label{fig:bib:n-100-m-100}
  \includegraphics[width=0.3\textwidth]{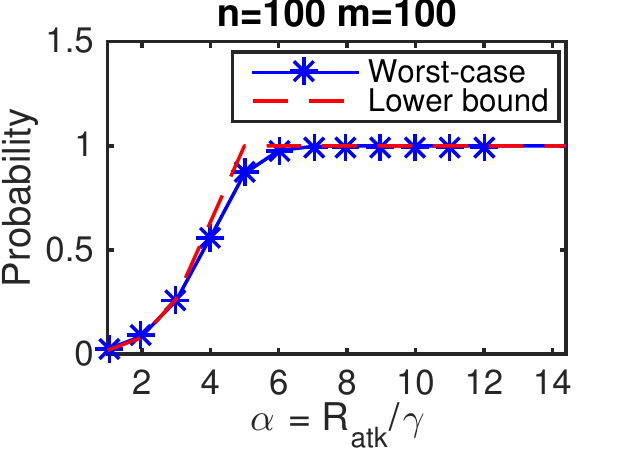}}
\subfigure[$n = 10^3$, $m = 100$] {
  \label{fig:bib:n-1000-m-100}
  \includegraphics[width=0.3\textwidth]{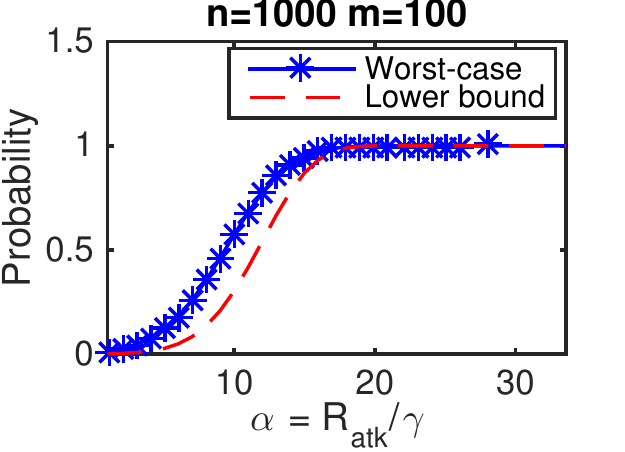}}
\subfigure[$n = 10^4$, $m = 100$] {
  \label{fig:bib:n-10000-m-100}
  \includegraphics[width=0.3\textwidth]{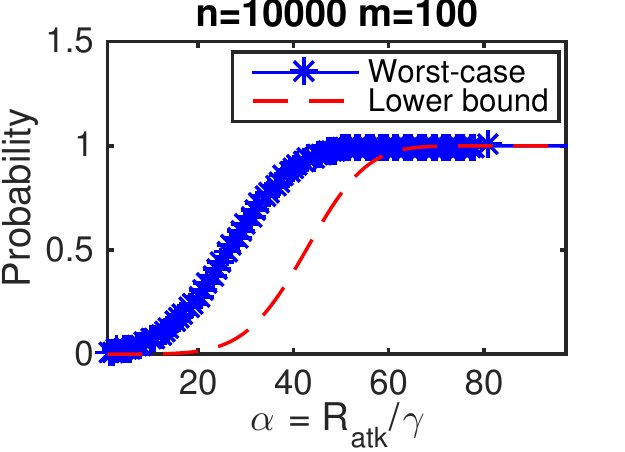}}
\subfigure[$n = 10^5$, $m = 100$] {
  \label{fig:bib:n-100000-m-100}
  \includegraphics[width=0.3\textwidth]{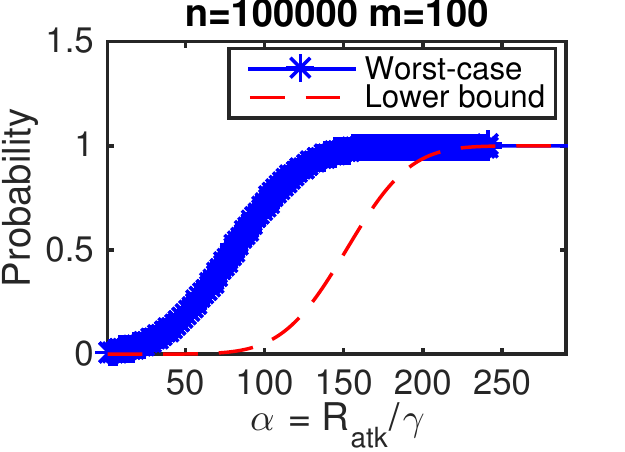}}
\subfigure[$n = 50$, $m = 1000$] {
  \label{fig:bib:n-50-m-1000}
  \includegraphics[width=0.3\textwidth]{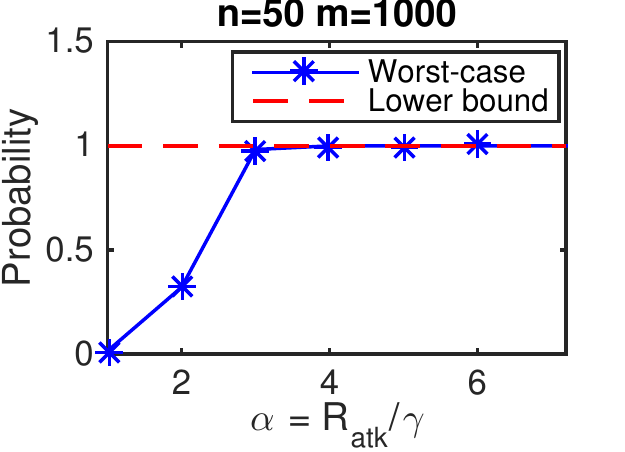}}
\subfigure[$n = 100$, $m = 1000$] {
  \label{fig:bib:n-100-m-1000}
  \includegraphics[width=0.3\textwidth]{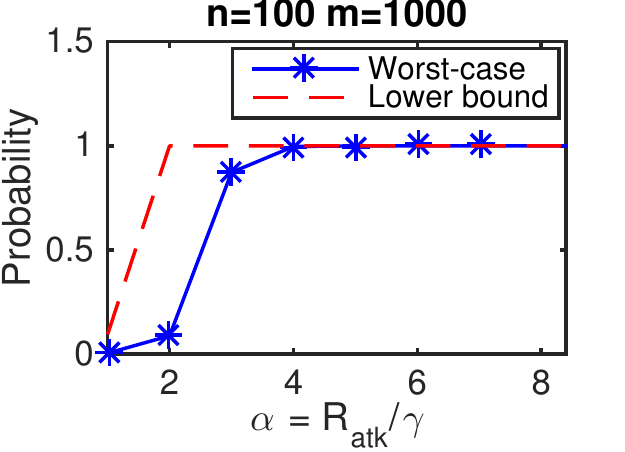}}
\subfigure[$n = 10^3$, $m = 1000$] {
  \label{fig:bib:n-1000-m-1000}
  \includegraphics[width=0.3\textwidth]{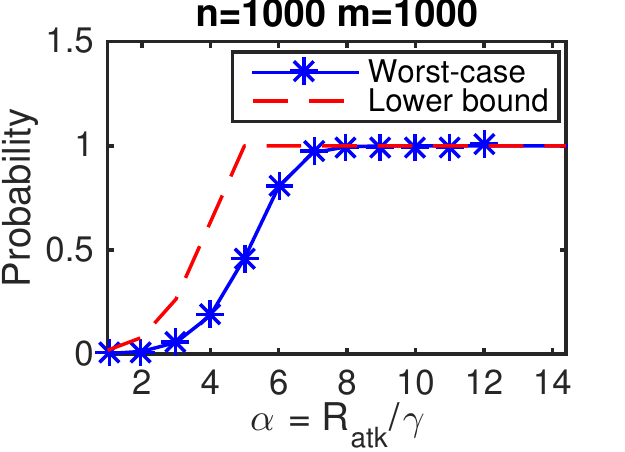}}
\subfigure[$n = 10^4$, $m = 1000$] {
  \label{fig:bib:n-10000-m-1000}
  \includegraphics[width=0.3\textwidth]{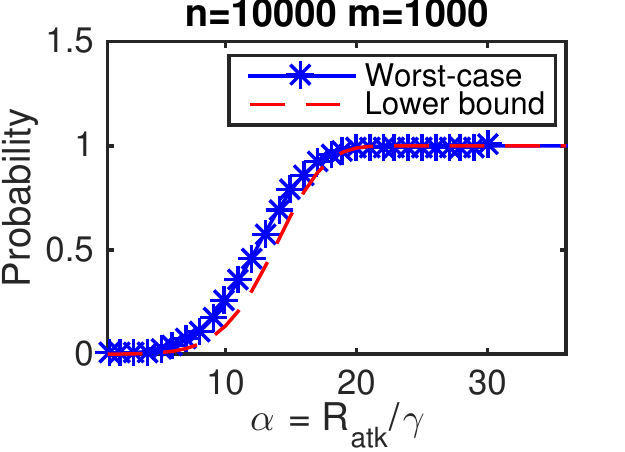}}
\subfigure[$n = 10^5$, $m = 1000$] {
  \label{fig:bib:n-100000-m-1000}
  \includegraphics[width=0.3\textwidth]{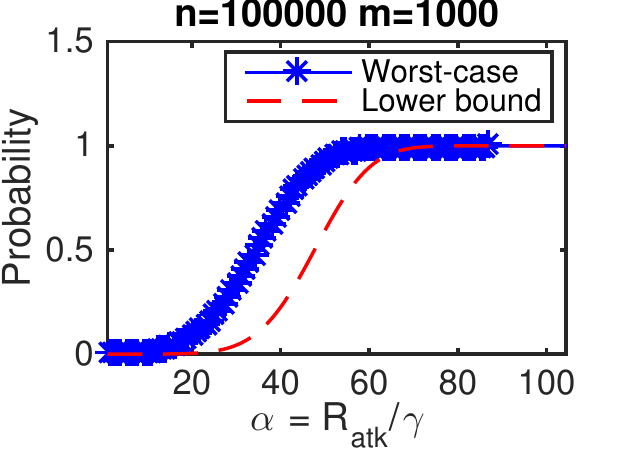}}
\caption{ The probability $\func{P_{\fmit{worst}}}{m, n, \alpha}$ when
$n$ full-use legitimate flows (at rate of $\gamma$), 
$m$ counters, and a large flow at the rate of 
$R_{\fmit{atk}} = \alpha\cdot \gamma$.}
\end{figure*}

\begin{figure*}[!th]
\centering
\subfigure[$\theta = 1.0$ (Flat)] {
  \label{fig:efd-td-flat}
  \includegraphics[width=0.3\textwidth]{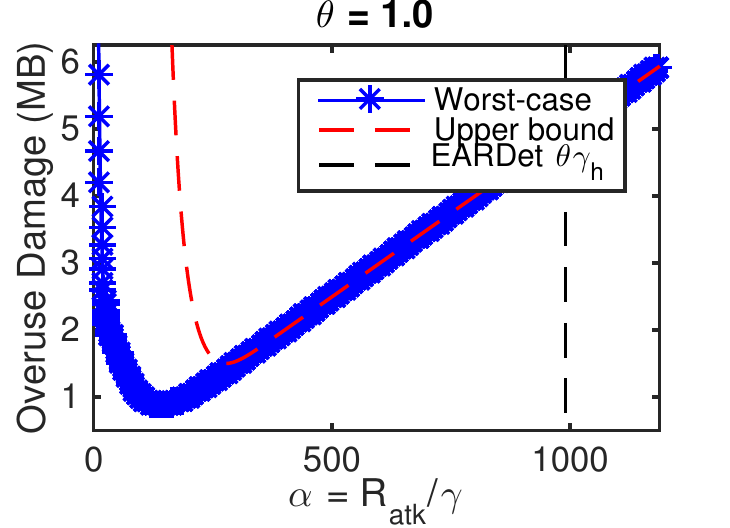}}
\subfigure[$\theta = 0.5$, $\theta T_b \ge 2T_c^{(1)}$] {
  \label{fig:efd-td-theta-05}
  \includegraphics[width=0.3\textwidth]{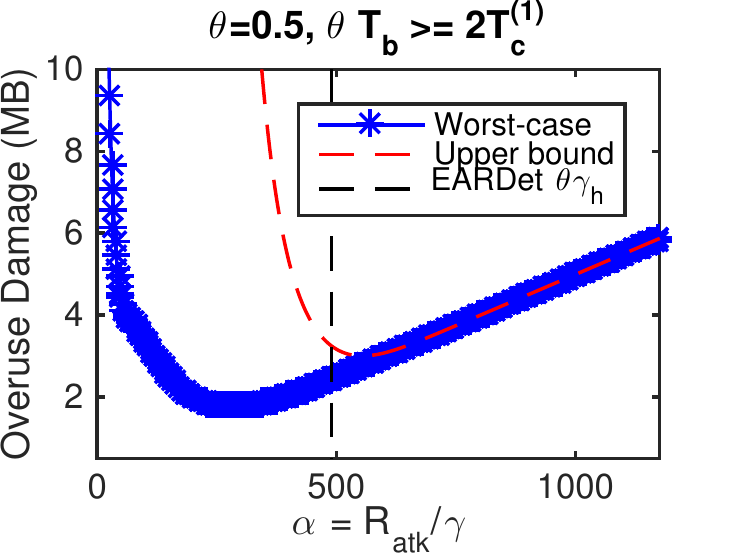}}
\subfigure[$\theta = .25$, $\theta T_b \ge 2T_c^{(1)}$] {
  \label{fig:efd-td-theta-025}
  \includegraphics[width=0.3\textwidth]{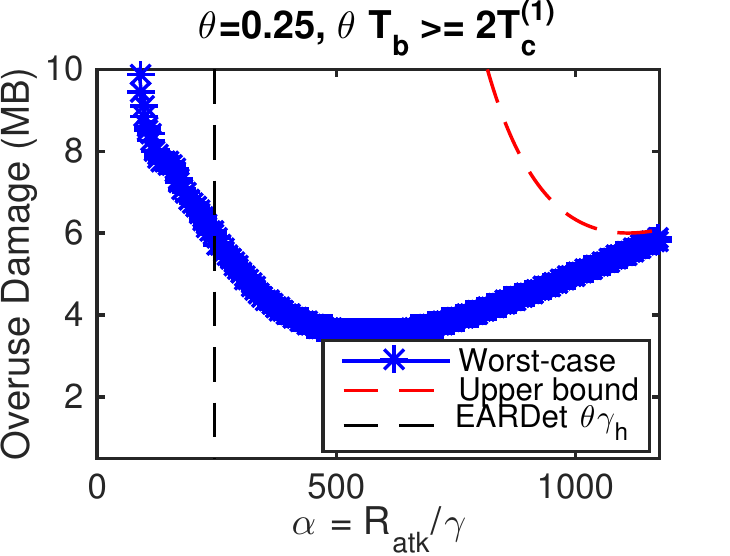}}
\subfigure[$\theta = 0.5$, $\theta T_b < 2T_c^{(1)}$] {
  \label{fig:efd-td-theta-05-twin-efd}
  \includegraphics[width=0.3\textwidth]{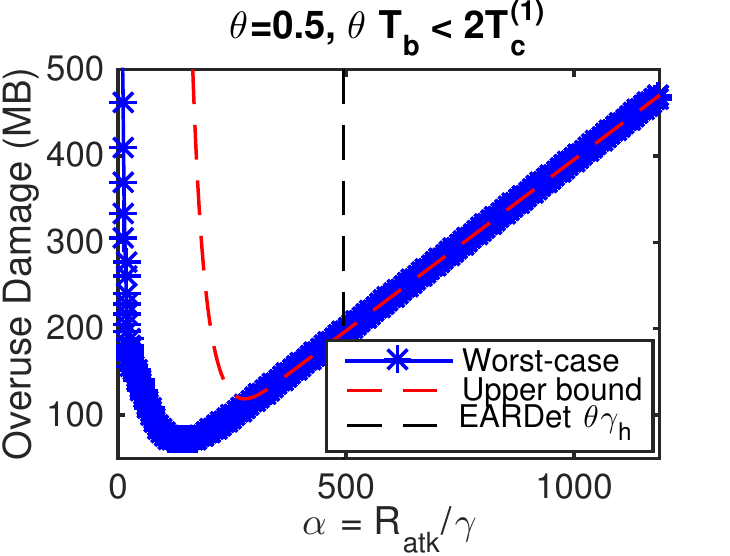}}
\subfigure[$\theta = .25$, $\theta T_b < 2T_c^{(1)}$] {
  \label{fig:efd-td-theta-025-twin-efd}
  \includegraphics[width=0.3\textwidth]{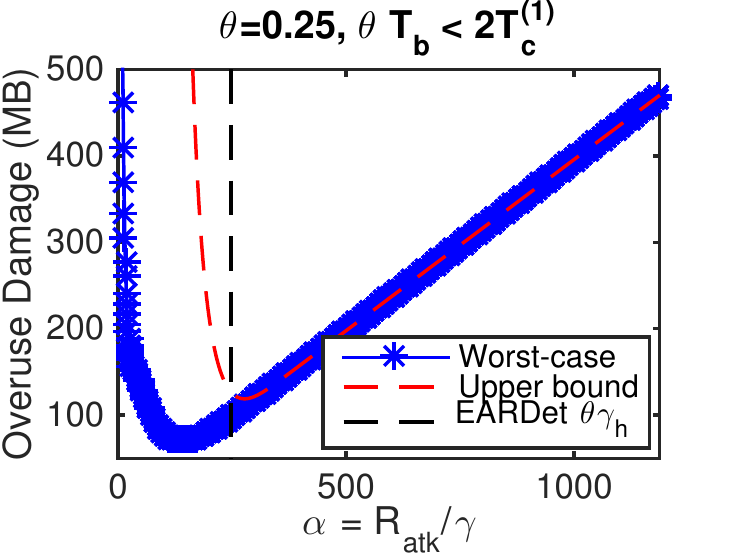}}
\caption{\twinEFD worst-case expected overuse damage $E(D_{\fmit{over}})$
(in MBytes) and its upper bound
for flat/bursty flows in various duty cycles $\theta$, 
burst periods $T_b$, and average rates $R_{\fmit{atk}} = \alpha\gamma$,
in the $40$ Gbps link with threshold rate $\gamma = 400$ Kbps 
($n_{\gamma} = 10^5$ full-use legitimate flows at most).
The \twinEFD has a limited memory of $m=100$ counters 
($50$ counters for each \EFD), 
a typical number of levels $d = 4$, and detection cycle $T_c^{(1)} = 0.1$ sec,
$T_c^{(2)}=7.92$ sec for two \EFDs respectively. 
Flows at the EARDet detectable rate
$R_{\fmit{atk}} \ge \theta\gamma_h = \theta\frac{n_{\gamma}}{m+1}\gamma$
are detected by the EARDet with $m=100$ counters in nearly zero damage.
}
\end{figure*}

\paragraph{Numeric analysis for single-level detection.}
For each theoretical result, we show numeric examples in the scenario of
$n_{\gamma} = 10^5$ and $m=100$, a even more memory-limited 
setting than the one in the complexity analysis 
(Section~\ref{ssec:efd-complexity}).

Figures~\ref{fig:bib:n-50-m-100} to~\ref{fig:bib:n-100000-m-1000}
comprehensively shows the simulated worst-case detection probability
$\func{P_{\fmit{worst}}}{m, n, \alpha}$ in a level
and its lower bound for various number of full-use legitimate flows 
$n \le n_{\gamma}$ ($50$ to $10^5$). 
We also give the numeric results with different 
$m$ ($100$ and $1000$) for comparison.
When $n=10^5,m=100$, we can see $\alpha_{0.5}=152$ and $\alpha_{1.0}=303$,
which are far smaller than EARDet's lowest detectable 
$\alpha=\frac{\rho}{\gamma(m+1)}=\frac{n_{\gamma}}{m+1}=991$.
For \EFD, the $\alpha$ with actual worst-case detection probability of 
$0.5$ and $1.0$ are around $75$ and $150$, respectively, 
which are much lower than the $\alpha_{0.5}$ and $\alpha_{1.0}$.
Thus, it suggests \EFD's ability of detecting low-rate large flows.
The figures also show that the probability of detecting low-rate flows increases as the number of flows ($n$) decreases or the number of counters ($m$) increases. 
The figures show that the $\alpha_{0.5}$, $\alpha_{1.0}$, 
and the lower bound holds for $n > m$, because 
we derive it with the assumption of $n \gg m\log m$.
When $n \le m$, \EFD has $100\%$ detection probability as explained in
Section~\ref{sec:hybrid},

Since $\alpha_{1.0}$ decreases rapidly when $n$ decreases, we approximate the
total detection probability by the detection probability of the first few levels.

\paragraph{Numeric analysis for total detection probability.}
In a tough scenario with $n_{\gamma} = 10^5$, $m=100$, 
and $n=10^7$ legitimate flows in a link during one detection cycle 
(around a second), \EFD has at least $0.25$ and $1.0$ probability 
to detect a flat large flow with $\alpha = 152$ and $303$, respectively;
and the simulation results suggest that \EFD can detect a large flow with
$\alpha = 75$ and $150$ with probability around $0.25$ and $1.0$, respectively.
Again, EARDet can only guarantee to detect 
$\alpha\ge\frac{\rho}{\gamma(m+1)}=\frac{n_{\gamma}}{m+1}=991$.
That is, \EFD outperforms the exact detection algorithm on low-rate large flows.

\begin{figure}[!th]
\centering
\includegraphics[width=0.8\textwidth]{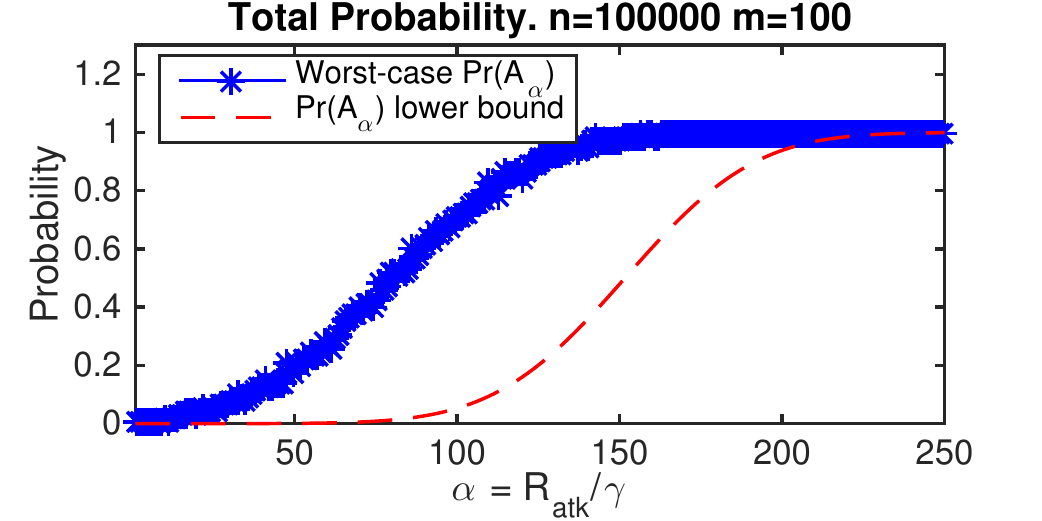}
\caption{\EFD Total Detection Probability. $n_{\gamma} = n = 10^5$, $m = 100$.}
\label{fig:efd-total-prob}
\end{figure}

Figure~\ref{fig:efd-total-prob} shows an example of simulated
worst-case total detection probability $Pr(A_{\alpha})$ and its
theoretical lower bound (Theorem~\ref{th:total-lower-bound}),
when $n=n_{\gamma}=10^5$ and $m=100$. The lower bound holds for the most
of $\alpha$, except some very small ones whose $Pr(A_{\alpha})$ is
close to $0$.

\subsection{Numeric Analysis For \twinEFD Theoretical Overuse Damage}
\label{ssec:appendix-efd-damage}
Figures~\ref{fig:efd-td-flat} to~\ref{fig:efd-td-theta-025-twin-efd}
show the expected overuse damage $E(D_{\fmit{over}})$
calculated in the worst case and its upper bound 
from Theorem~\ref{th:twin-efd-bursty-damage}, in a $40$ Gbps link with
threshold rate $\gamma=400$ Kbps ($n_{\gamma} = 10^5$ full-use legitimate
flows at most). The \twinEFD has $m=50$ counters for each \EFD, 
$d=4$ levels, and detection cycle $T_c^{(1)} = 0.1$ sec and $T_c^{(2)} = 7.92$
sec for two \EFDs, respectively. 
Damages by large flows with various duty cycles $\theta$ 
and burst periods $T_b$ are shown.
Flows with an average rate $R_{\fmit{atk}}$ higher than $\theta\gamma_h$ (black dash line)
will be detected instantly by EARDet (with $100$ counters) with nearly zero damage.

The \twinEFD has $d=4$ so that
the number of virtual counters in the \EFD bottom level
($m^d = 50^4 = 6.25\times10^6$) is larger than the number of flows.
Therefore, we the flows selected to the bottom level is fewer than
the counters, and \EFD can track each flow individually in the bottom level.

We set $T_c^{(1)}=T_c=0.1$ sec around 
$\frac{\beta}{\gamma}=\frac{2\times1514}{400Kbps}=0.06$ sec 
($\beta$ is usually a few times of maximum packet size $1514$ Bytes,
so that bursty flows are easier to catch).
If $T_c \ll \beta/\gamma$, 
it is hard for a large flow to reach burst threshold;
if $T_c \gg \beta/\gamma$, the detection delay is too long,
resulting in excessive damage.

For \twinEFD's second \EFD,
$T_c^{(2)} = \frac{2d\gamma_h}{\alpha\gamma}T_c^{(1)} = 7.92$ sec
(according to Theorem~\ref{th:twin-efd-bursty-damage}).
Therefore, \twinEFD can guarantee detection for 
the worst-case bursty flows ($\theta T_b < 2T_c^{(1)}$) at rate
$R_{\fmit{atk}} \ge \alpha\gamma = 100\gamma$.
We can guarantee detection of lower rate flows with worst-case burstiness
by increasing $T_c^{(2)}$; however, increased $T_c^{(2)}$ increases
the damage caused by worst-case bursty flows.
We say bursty flows with
$\theta T_b < 2T_c^{(1)}$ are the worst-case bursty flows, because such flows
are unlikely showing up in every level of the \EFD with cycle ($T_c^{(1)}$),
so that we have to use the \EFD with longer detection cycle ($T_c^{(2)}$) to
catch those flows, which requires longer delay, thus higher damage.
Furthermore, such worst-case flows can inflict more damage by increasing
$\theta$ (thus the average rate), but remain undetectable by EARDet.
As discussed in Section~\ref{ssec:eardet-efd-hybrid},
we can use choose different $T_c^{(2)}$ randomly in different cycles 
to prevent attackers from deterministically maximizing damage.

A hybrid scheme consisting of EARDet and \twinEFD can limit the worst-case
damage caused by flat flows ($\theta=1$) and bursty flows ($\theta <
1$). Specifically, for flows with an average rate larger than $30\gamma$
(i.e. $12$ Mbps), the damage is as low as tens of MBytes (less than ten MBytes
for flat flows). 
We admit that \twinEFD cannot limit the damage for 
flows at extremely low rate ($\ll 30\gamma$) as effectively as for other flows,
however other existing schemes cannot neither, because of the limited memory.
For flows at high rates, although the \twinEFD detects them with
almost $100\%$ probability in one detect cycle, it requires at least
one cycle to finish detection, 
hence the damage increases linearly with the flow rate.

\twinEFD and EARDet complement each other. \twinEFD can detect flows with an
average rate lower than $\theta\gamma_h$ but it incompetent at detecting
high-rate flows, whereas EARDet is the opposite.

\section{Proof Sketches}
\subsection{Proof Sketch For \EFD Single-level Detection Probability}
\label{ssec:appendix-efd-single-level-prob-proof}

\paragraph{Proof sketch of Theorem~\ref{th:efd-level-prob}}
In the analysis, we treat hashing flows into counters 
as uniformly assigning $n$ legitimate flows
into counters and pick a counter for 
the large flow $f_{\fmit{atk}}$ at random.
We denote the random variable of the maximum number of legitimate flows
assigned to a counter as $Y$ and the random variable of 
the number of legitimate flows in the counter of 
the large flow $f_{\fmit{atk}}$ as $X$. 

Because the \EFD pick the counter with the largest value for 
the next level, thus as long as the value of the large-flow counter 
$(R_{\fmit{atk}} + X \cdot \gamma)T$ is higher than the value of the
maximum-value legitimate counter $Y\cdot \gamma T$, the large-flow counter 
will be picked, where $T$ is the time length of the level. Then we get

\begin{small}
\begin{equation}\label{eq:detect-prob}
    \begin{aligned}
        \func{P_{\fmit{worst}}}{m, n, \alpha}
        =& Pr(R_{\fmit{atk}} + X \cdot \gamma - Y \cdot \gamma > 0)\\
        =& Pr(Y - X < \alpha)\\
        =& \sum_{y} Pr(y - X < \alpha | Y = y) \cdot Pr(Y = y)
    \end{aligned}
\end{equation}
\end{small}

As we discussed in Section~\ref{ssec:appendix-worst-efd},
the distributions of $X$ and $Y$ are the same as those of
the $X_b$ and $Y_b$ in a single-weight balls-into-bins game 
with $n$ balls and $m$ bins, where the $X_b$ is the random variable of the
number of balls in a randomly picked bin, and the $Y_b$ is the random variables
of the maximum number of balls in a counter.
Thus, we can apply Theorem~\ref{th:max-load} 
(by Raab and Steger~\cite{raab1998balls}) to calculate
the $y_{\fmit{max}}$, the upper bound of $Y$ at high probability.

\begin{theorem}\label{th:max-load}
\textbf{Raab and Steger's Theorem 1.} 
Let $Y$ be the random variable that counts 
the maximum number of balls in any bin,
if we throw $n$ balls independently and uniformly at random into $m$ bins.
Then $Pr(Y > y_{\fmit{max}}) = o(1)$, 
if $y_{\fmit{max}} = \frac{n}{m} + \lambda\sqrt{2\frac{n}{m}\log n}$, 
$\lambda > 1$, $m\log m \ll n \le m \cdot ploylog(m)$, and $n$ is very large.
When $n \rightarrow \infty$, $o(1) \rightarrow 0$.
\end{theorem} 

We think it is a good approximation to our large-flow problem.
Because the number of legitimate flows $n$ 
in a backbone link is more than a million,
while the number of counters $m$ is quite limited 
(e.g. one thousand counters in L1 cache),
thus we say $m\log m \ll n \le m \cdot ploylog(m)$
\footnote{Raab~\etal also provides a similar $y_{\fmit{max}}$ for 
$m(\log m)^3 \ll n$, but it is enough to only discuss one of them
for an approximate result.}
and $n$ is very large. We derive the approximate lower bound of 
$\func{P_{\fmit{worst}}}{m, n, \alpha}$ as follows:

\begin{small}
\begin{equation}\label{eq:detect-prob-bound-1}
    \begin{aligned}
        &\func{P_{\fmit{worst}}}{m, n, \alpha} 
        = \sum_{y} Pr(y - X < \alpha | Y = y) \cdot Pr(Y = y) \\
        &= \sum_{y \le y_{\ffmit{max}}} Pr(X>y-\alpha, Y=y)
            + \sum_{y > y_{\ffmit{max}}} Pr(X>y-\alpha, Y=y)\\
        &\ge \sum_{y \le y_{\ffmit{max}}} Pr(X>y_{\fmit{max}}-\alpha, Y=y)
            + \sum_{y > y_{\ffmit{max}}} Pr(X>y-\alpha, Y=y)\\
        &= Pr(X>y_{\fmit{max}}-\alpha)\cdot Pr(Y \le y_{\fmit{max}})
            + \sum_{y > y_{\ffmit{max}}} Pr(X>y-\alpha, Y=y)
    \end{aligned}
\end{equation}
\end{small}

\noindent 
We prove that the second part is $o(1)$ as follows,
\begin{small}
\begin{equation}\label{eq:detect-prob-bound-2}
    \begin{aligned}
        \sum_{y > y_{\ffmit{max}}} Pr(X>y-\alpha, Y=y)
        &\le \sum_{y > y_{\ffmit{max}}} Pr(X>y_{\fmit{max}}-\alpha, Y=y) \\
        &= Pr(X > y_{\fmit{max}-\alpha})\cdot Pr(Y>y_{\fmit{max}-\alpha}) \\
        &= Pr(X > y_{\fmit{max}-\alpha})\cdot o(1) = o(1)
    \end{aligned}
\end{equation}
\end{small}

\noindent
According to Equation~\ref{eq:detect-prob-bound-1} 
and~\ref{eq:detect-prob-bound-2}, we get

\begin{small}
\begin{equation}\label{eq:detect-prob-bound-3}
    \begin{aligned}
        \func{P_{\fmit{worst}}}{m, n, \alpha}  
        &\ge Pr(X>y_{\fmit{max}}-\alpha)\cdot 
            Pr(Y \le y_{\fmit{max}})+o(1) \\
        &=Pr(X>y_{\fmit{max}}-\alpha)\cdot 
            [1 - Pr(Y > y_{\fmit{max}})] + o(1) \\
        &=Pr(X>y_{\fmit{max}}-\alpha)\cdot (1 - o(1)) + o(1) \\
        &=Pr(X>y_{\fmit{max}}-\alpha) - o(1) \\
        &\approx Pr(X>y_{\fmit{max}}-\alpha)
    \end{aligned}
\end{equation}
\end{small}

\noindent
Therefore, when $n$ is large, we approximately have
$\func{P_{\fmit{worst}}}{m, n, \alpha} \ge Pr(X>y_{\fmit{max}}-\alpha)$.

We let $\eta = \lceil y_{\fmit{max}}-\alpha \rceil$, 
and use random variable $M_k$ 
to denote the number of bins exactly contain $k$ balls.
We calculate $Pr(X>y_{\fmit{max}}-\alpha)$ as follows,

\begin{small}
\begin{equation}\label{eq:detect-prob-bound-4}
    \begin{aligned}
        &Pr(X>y_{\fmit{max}}-\alpha)\\
        &= Pr(X>\eta) = \sum_{k \ge \eta} Pr(X = k) \\
        &= \sum_{k \ge \eta} \sum_{0 \le m_k \le m} Pr(X = k | M_k = m_k) 
            \cdot Pr(M_k = m_k) \\
        &= \sum_{k \ge \eta} \sum_{0 \le m_k \le m} 
            \frac{m_k}{m} Pr(M_k = m_k) \\
        &= \sum_{k \ge \eta} \frac{1}{m} 
            \sum_{0 \le m_k \le m} m_k Pr(M_k = m_k)
            = \sum_{k \ge \eta} \frac{E(M_k)}{m} \\
        &= \sum_{\eta \le k \le n} \frac{1}{m} \cdot 
            m\frac{\binom{n}{k}(m-1)^{n-k}}{m^n} \\
        &= \sum_{\eta \le k \le n} \binom{n}{k} 
            \bigg(1 - \frac{1}{m} \bigg)^{n-k} \bigg(\frac{1}{m} \bigg)^k
    \end{aligned}
\end{equation}
\end{small}

The above result requires to calculate 
the sum of the last $m - \eta + 1$ items from the binomial distribution
$B(n, \frac{1}{m})$. As we know, there is no simple, 
closed forms for Equation~\ref{eq:detect-prob-bound-4}.
According to the law of rare events~\cite{cameron2013regression},
binomial distribution $B(n, p)$ is approximate to Poisson distribution
$\mit{Pois}(np)$, when $n$ is large and $p$ is small.
According to Equation~\ref{eq:total-detect-prob-3},
the detection probability $Pr(A_{\alpha})$ is mainly related to 
non-bottom levels in which the number of flows $n$is large ($n > m$)
, and $p = \frac{1}{m}$ is small because $m$ is around hundreds
to thousands, we approximately treat $B(n, \frac{1}{m})$ as 
the Poisson distribution $\mit{Pois}(\frac{n}{m})$, then we have

\begin{small}
\begin{equation}\label{eq:binomial-to-poisson}
    \binom{n}{k} \bigg(1 - \frac{1}{m} \bigg)^{n-k} \bigg(\frac{1}{m} \bigg)^k
    \approx \frac{e^{-\frac{n}{m}}(\frac{n}{m})^k}{k!}
\end{equation}
\end{small}

\noindent
, which is the probability of the item happens $k$ times 
in the Poisson distribution.
Then, the Equation~\ref{eq:detect-prob-bound-4} turns to

\begin{small}
\begin{equation}\label{eq:detect-prob-bound-5}
    \begin{aligned}
        Pr(X>y_{\fmit{max}}-\alpha) 
        &= \sum_{\eta \le k \le n} \binom{n}{k} 
            \bigg(1 - \frac{1}{m} \bigg)^{n-k} \bigg(\frac{1}{m} \bigg)^k\\
        &= \sum_{\eta \le k \le n} \frac{e^{-\frac{n}{m}}(\frac{n}{m})^k}{k!}
         = 1 - Q(\eta-1, \frac{n}{m})
    \end{aligned}
\end{equation}
\end{small}

\noindent
, where $Q(K, \frac{n}{m})$ is the cumulative distribution function (CDF)
of the Poisson distribution $\mit{Pois}(\frac{n}{m})$, 
i.e. sum of probabilities for $0 \le k \le K $.
As the Theorem~\ref{th:max-load} holds when $\lambda > 1$, 
thus we choose $\lambda \rightarrow 1^+$, 
thus $y_{\fmit{max}} = \frac{n}{m}+\sqrt{2\frac{n}{m}\log n}$. 
Because we focus on how does the probability lower bound change 
along with the $m$ and $n$, the $\lambda$ does not matter much here.
Therefore, we proved that the $1 - Q(K, \frac{n}{m})$ is 
an approximate lower bound for $\func{P_{\fmit{worst}}}{m, n, \alpha}$, where 
$K=\eta-1=\bigl\lfloor \frac{n}{m}+\sqrt{2\frac{n}{m}\log n} - \alpha \bigr\rfloor$.
$\blacksquare$

\paragraph{Proof sketch of Corollary~\ref{cl:0.5-lower-bound}.}
According to Theorem~\ref{th:efd-level-prob},
$\func{P_{\fmit{worst}}}{m, n, \alpha_{0.5}} > 1 - Q(K, \frac{n}{m})$
approximately, where 
$K = \bigl\lfloor \frac{n}{m}+\sqrt{2\frac{n}{m}\log n} - \alpha_{0.5} \bigr\rfloor$. 
As the median\footnote{The $K$ such that the CDF $Q(K, \frac{n}{m}) = 0.5$} 
$\nu$ of the Poisson distribution $\mit{Pois}(\frac{n}{m})$
is bounded by $\frac{n}{m} - \log 2 \le \nu < \frac{n}{m} + \frac{1}{3}$
~\cite{choi1994medians}. Thus, $\nu \approx \frac{n}{m}$, then

\begin{small}
\begin{equation}\label{eq:r_atk_50}
    K \approx \frac{n}{m} \Rightarrow 
        \alpha_{0.5} \approx \sqrt{2\frac{n}{m}\log n}
\end{equation}
\end{small}
\noindent
Therefore the Corollary~\ref{cl:0.5-lower-bound} is proved.$\blacksquare$

\paragraph{Proof sketch of Corollary~\ref{cl:1.0-lower-bound}.}
According to Pearson's Skewness Coefficients~\cite{skewness}, the symmetry of
a distribution is measured by its skewness. The probability distribution 
is approximately symmetrical to its mean when the skewness is small.
According to~\cite{poisson-distribution}, 
the skewness of Poisson distribution $\mit{Pois}(\frac{n}{m})$
is $\big(\frac{n}{m}\big)^{-0.5}$. Thus when $n \gg m\log m$ the 
$Pois(\frac{n}{m})$ is approximately symmetrical to its mean $\frac{n}{m}$.

Because when $\alpha = 1$ the actual $\func{P_{\fmit{worst}}}{m, n, \alpha}$
should be $\frac{1}{m} \approx 0$ (because the large flow rate is the same
as the legitimate flow rate, thus the detection equals to randomly 
picking one from $m$ counters), thus the approximate lower bound 
$1-Q(K_{\alpha=1}, \frac{n}{m}) \approx 0$, where 
$K_{\alpha=1} = \bigl\lfloor \frac{n}{m} + \sqrt{2\frac{n}{m}\log n} - 1\bigr\rfloor$.
As $\mit{Pois}(\frac{n}{m})$ is symmetrical to $K_s = \frac{n}{m}$,
when $K = K_s+(K_s-K_{\alpha=1}) \approx \frac{n}{m}-\sqrt{2\frac{n}{m}\log n}$
, the $1 - Q(K, \frac{n}{m}) \approx 1$, 
in which $\alpha \approx 2\alpha_{0.5}$ 
(according to Corollary~\ref{cl:0.5-lower-bound}). Thus,
$\alpha_{1.0} = 2\alpha_{0.5}$ has been proved.$\blacksquare$

\subsection{Proof Sketch For \EFD Total Detection Probability}
\label{ssec:appendix-efd-total-prob-proof}
\paragraph{Proof sketch of Theorem~\ref{th:total-lower-bound}.}
For the detection level $k$, we use $A_{k,\alpha}$ to denote 
the event that the counter containing the large flow $f_{\fmit{atk}}$ 
with average rate of $R_{\fmit{atk}}=\alpha\gamma$ in the level $k$
is selected for the next level, where $\gamma$ is the threshold rate, 
$\alpha > 1$. Then the total probability for \EFD to catch
the large flow $f_{\fmit{atk}}$ in one detection cycle is

\begin{small}
\begin{equation}\label{eq:total-detect-prob}
    \begin{aligned}
        Pr(A_{\alpha}) =& Pr(A_{1,\alpha}, A_{2,\alpha}, A_{3,\alpha}
            ,...,A_{d,\alpha}) \\
        =& Pr(A_{1,\alpha})\cdot Pr(A_{2,\alpha}|A_{1,\alpha})\cdot
            Pr(A_{3,\alpha}|A_{2,\alpha},A_{1,\alpha})\cdot \\         
            &...Pr(A_{d,\alpha}|A_{d-1,\alpha},...,A_{1,\alpha}) \\
        =& Pr(A_{1,\alpha})\cdot Pr(A_{2,\alpha}|A_{1,\alpha})\cdot
            Pr(A_{3,\alpha}|A_{2,\alpha})\cdot \\         
            &...Pr(A_{d,\alpha}|A_{d-1,\alpha})
    \end{aligned}
\end{equation}
\end{small}

As we described in Section~\ref{ssec:efd-optimizations}, 
we use the Cuckoo hashing in the bottom level $d$ to randomly assign flows into
counters. Because we set enough levels to make 
the input flows in the bottom level less than the counters, 
the $Pr(A_{d,\alpha}|A_{d-1,\alpha}) \approx 1$. For the levels $k < d$
with $n^{(k)}$ legitimate flows, according to Theorem~\ref{th:efd-level-prob}
the $Pr(A_{k,\alpha}|A_{k-1,\alpha}) \ge
\func{P_{\fmit{worst}}}{m, n^{(k)}, \alpha}$.
Considering the maximum number of full-use legitimate flows in a link
is $n_{\gamma} = \rho/\gamma$,

\begin{itemize}
\item When $n < n_{\gamma}$, $Pr(A_{k,\alpha}|A_{k-1,\alpha}) \ge 
\func{P_{\fmit{worst}}}{m, n^{(k)}, \alpha}$

\item When $n \ge n_{\gamma}$, $Pr(A_{k,\alpha}|A_{k-1,\alpha}) \ge 
\func{P_{\fmit{worst}}}{m, n_{\gamma}, \alpha}$
\end{itemize}

\noindent
Therefore,
\begin{small}
\begin{equation}\label{eq:total-detect-prob-3}
    \begin{aligned}
        Pr(A_{\alpha}) \ge \prod^{d-1}_{k=1} \func{P_{\fmit{worst}}}{m, 
                \func{min}{n_{\gamma}, n^{(k)}}, \alpha}
    \end{aligned}
\end{equation}
\end{small}
\noindent
, where we approximately let $n^{(k)} = n/m^{k-1}$, which is the average
value of $n^{(k)}$ over repeated detection. $n$ is the number of 
legitimate flows in the link.

According to Equation~\ref{eq:total-detect-prob-3} and the fact that 
$\alpha_{1.0}$ decreases fast as the $n^{(k)}$ decreases by the factor of $m$,
$\func{P_{\fmit{worst}}}{m,n^{(k)},\alpha}$ for $n^{(k)} < n_{\gamma}$ 
does not affect the product much for the most of $\alpha$ values. 
Therefore, we can approximate $Pr_{\fmit{worst}}(A_{\alpha})$ as follows:

\begin{small}
\begin{equation}\label{eq:approx-worst-total-prob}
Pr(A_{\alpha}) \ge \left\{ \begin{aligned}
    &\prod_{\{k| n^{(k)} \ge n_{\gamma}\}} \func{P_{\fmit{worst}}}{m,n_{\gamma},\alpha}
        \mit{, when $n \ge n_{\gamma}$}\\
    &\func{P_{\fmit{worst}}}{m,n,\alpha}
        \hspace{10 mm} \mit{, when $n < n_{\gamma}$}\\
\end{aligned} \right.
\end{equation}
\end{small}

\noindent
, where size of $\{k| n^{(k)} \ge n_{\gamma}\}$ is 
$\lfloor \log_m (n/n_{\gamma}) \rfloor + 1$, 
because $n^{(k)} = n^{(k-1)}/m$.

According to Theorem~\ref{th:efd-level-prob}, approximately
$\func{P_{\fmit{worst}}}{m,n,\alpha} \ge 1 - Q(K, \frac{n}{m})$
where $K = \bigl\lfloor \frac{n}{m} + \sqrt{2\frac{n}{m}\log n} - \alpha \bigr\rfloor$.
Thus we can derive Theorem~\ref{th:total-lower-bound}
from Equation~\ref{eq:approx-worst-total-prob}.$\blacksquare$

\subsection{Proof Sketch For \twinEFD Theoretical Overuse Damage}
\label{ssec:appendix-efd-damage-proof}

The upper bound of the expected overuse damage can be derived 
from the average rate of a flat large flow and the expected detection delay: 
$E(D_{\fmit{over}}) \le E(T_{\fmit{delay}})\cdot R_{\fmit{atk}}$, 
because attack flows cannot cause more overuse damage than 
the amount of traffic over-sent $E(T_{\fmit{delay}})\cdot R_{\fmit{atk}}$.
For a bursty flow with duty cycle
$\theta$ and burst period $T_b$, a \EFD can also treat it as a flat flow at the
time of each burst interval $\theta T_b$. Thus, we can still use the detection
probability for flat flows to calculate the damage for bursty flows.

\paragraph{Lemma~\ref{lm:bursty-damage-upper-bound} and proof sketch.}
\begin{lemma}\label{lm:bursty-damage-upper-bound}
A \EFD with detection cycle $T_c$ can detect bursty flows with 
$\theta T_b \ge 2T_c$ with the expected overuse damage:

\begin{small}
\begin{equation}\label{eq:bursty-damage-upper-bound}
E(D_{\fmit{over}}) \le \left\{ \begin{aligned}
&T_{c}\gamma\alpha / \theta \bigg(1 - Q(K_{\gamma}, \frac{n_{\gamma}}{m})\bigg)
    ^{\lfloor \log_m (n/n_{\gamma}) \rfloor + 1}
    \mit{, when $n \ge n_{\gamma}$}\\
&T_{c}\gamma\alpha / \theta \bigg(1 - Q(K, \frac{n}{m}))\bigg) \hspace{15 mm} \mit{, when $n < n_{\gamma}$}\\
\end{aligned} \right.
\end{equation}
\end{small}
\noindent
where $K_{\gamma} = \bigl\lfloor \frac{n_{\gamma}}{m} + \sqrt{2\frac{n_{\gamma}}{m}\log n_{\gamma}} - \frac{\alpha}{\theta} \bigr\rfloor$ 
, $K = \bigl\lfloor \frac{n}{m} + \sqrt{2\frac{n}{m}\log n} - \frac{\alpha}{\theta} \bigr\rfloor$,
and $Q(x,\lambda)$ is the CDF of the Poisson distribution $\mit{Pois}(\lambda)$.
\end{lemma}

\paragraph{Proof sketch:}
Because $\theta T_b \ge 2T_c$, thus for each burst period $T_b$ 
there are must be at least $\bigl\lfloor \frac{\theta T_b}{T_c} - 1 \bigr\rfloor$ 
detection cycles, in which \EFD can see the attack traffic in all levels.
When the \EFD observes the bursty flow, the only difference from the detection
over flat flow is that, the traffic rate at that moment is 
$\frac{\alpha}{\theta}\gamma$, instead of $\alpha\gamma$ 
in the case of flat flows.
Thus, the probability $Pr(A_{\alpha})$ 
to detect such bursty flow in one detection cycle
is calculated as the one for flat flow detection in 
Theorem~\ref{th:total-lower-bound}, by replacing the $\alpha$ with
the $\frac{\alpha}{\theta}$.

The expected detection delay $E(T_{\fmit{delay}})$ is derived as follows:
\begin{small}
\begin{equation}
E(T_{\fmit{delay}}) \le \frac{1}{Pr(A_{\alpha})}
    \frac{T_b}{\bigl\lfloor \frac{\theta T_b}{T_c} - 1 \bigr\rfloor}
    \approx \frac{T_c}{\theta Pr(A_{\alpha})}\\
\end{equation}
\end{small}
\noindent
Then we get the over-sent attack traffic in the input link is 
$E(T_{\fmit{delay}}) \cdot R_{\fmit{atk}}$, 
and the overused bandwidth by attack traffic is less than or equal to that, 
because the some attack packets may also be dropped during congestion.
Thus we get the expected overuse damage $E(D_{\fmit{over}})$:
\begin{small}
\begin{equation}
E(D_{\fmit{over}}) \le E(T_{\fmit{delay}}) \cdot R_{\fmit{atk}}
    \le T_c \gamma\alpha / \theta Pr(A_{\alpha})
\end{equation}
\end{small}
\noindent
Thus, according to Theorem~\ref{th:total-lower-bound}, we get the upper bound
of the overuse damage in the Lemma~\ref{lm:bursty-damage-upper-bound}. 
The proof also holds when $\theta = 1$, which is for the case of flat flows.
$\blacksquare$

\paragraph{Proof sketch of Theorem~\ref{th:twin-efd-bursty-damage}.}
The overuse damage in the case of $\theta T_b \ge 2T_c^{(1)}$ are from
Lemma~\ref{lm:bursty-damage-upper-bound}. 
When $\theta T_b < 2T_c^{(1)}$ and $R_{\fmit{atk}} < \theta\gamma_h$,
we prove the damage as follows:

\begin{small}
\begin{equation}
\begin{aligned}
&T_b < \frac{2T_c^{(1)}}{\theta} < \frac{2T_c^{(1)}}{R_{\fmit{atk}}}{\gamma_h}
    = \frac{2T_c^{(1)}}{\alpha\gamma}{\gamma_h} = \frac{T_c^{(2)}}{d}
\end{aligned}
\end{equation}
\end{small}
\noindent
Thus the $T_b$ is less than a detection level period of the $EFD^{(2)}$,
which means the bursty flow is like a flat flow to $EFD^{(2)}$.
Therefore, we use the overuse damage upper bound 
in Lemma~\ref{lm:bursty-damage-upper-bound},
when $\theta = 1$, $T_c = T_c^{(2)}$, and we get
\begin{small}
\begin{equation}
E(D_{\fmit{over}}) \le \left\{ \begin{aligned}
&T_{c}^{(2)}\gamma\alpha / \bigg(1 - Q(K_{\gamma}, \frac{n_{\gamma}}{m})\bigg)
    ^{\lfloor \log_m (n/n_{\gamma}) \rfloor + 1}
    \mit{, when $n \ge n_{\gamma}$}\\
&T_{c}^{(2)}\gamma\alpha / \bigg(1 - Q(K, \frac{n}{m}))\bigg) \hspace{15 mm} \mit{, when $n < n_{\gamma}$}\\
\end{aligned} \right.
\end{equation}
\end{small}
\noindent
where $K_{\gamma} = \bigl\lfloor \frac{n_{\gamma}}{m} + \sqrt{2\frac{n_{\gamma}}{m}\log n_{\gamma}} - \alpha \bigr\rfloor$ 
, $K = \bigl\lfloor \frac{n}{m} + \sqrt{2\frac{n}{m}\log n} - \alpha \bigr\rfloor$.
By replacing $T_c^{(2)}$ with $\frac{2\gamma_h}{\alpha\gamma}T_c^{(1)}$,
we proved the Theorem~\ref{eq:twin-efd-bursty-damage}.$\blacksquare$

\section{Additional Table}
\label{ssec:appendix-table-figure}

\begin{table}[htbp]\caption{Settings of \EFD and \ETE}
\label{table:efd-setting} 
\centering
\begin{small}
  \begin{tabular}{|c|c|c|c|c|c|c|c|}
    \hline
    \hline
    $m$ & $20$ & $40$ & $70$ & $100$ & $150$ & $200$ & $400$\\
    \hline
    $T_{\ell}^{*}$ & $.242$ & $.242$ & $.242$ & $.242$ & $.242$ & $.242$ 
    & $.242$\\
    \hline
    \multicolumn{8}{|c|}{Single \EFD} \\
    \hline
    $d$ & $4$ & $3$ & $3$ & $3$ & $3$ & $3$ & $2$\\
    \hline
    $T_c^{*}$ & $.968$ & $.726$ & $.726$ & $.726$ & $.726$ & $.726$ 
    & $.484$\\
    \hline
    \multicolumn{8}{|c|}{\twinEFD (in \ETE)} \\
    \hline
    $d$ & $7$ & $5$ & $4$ & $4$ & $4$ & $3$ & $3$\\
    \hline
    $T_c^{(1)*}$ & $1.69$ & $1.21$ & $.968$ & $.968$ & $.968$ & $.726$ 
    & $.726$\\
    \hline
    $T_c^{(2)*}$ & $168.6$ & $63.75$ & $31.92$ & $26.56$ & $21.96$ & $10.68$ 
    & $7.59$ \\
    \hline
  \end{tabular}

  $^{*}$ Time unit is second.

\end{small}
\end{table}

\fi

\end{document}